\documentclass[aip, onecolumn]{revtex4-2}
\usepackage{epstopdf}
\usepackage[caption=false]{subfig}
\usepackage{amsmath}
\usepackage{amssymb}

\newcommand \grad  {\mathbf{grad} \, }   
\newcommand \ddiv  {\mathrm{div} \, }   
\newcommand \curl  { \mathbf{curl} \;}   
\newcommand \Grad  {\mathbf{Grad} \, }   
\newcommand \Ddiv  {\mathrm{Div} \, }   
\newcommand \Curl  { \mathbf{Curl} \;}   

\begin{document}

\title{Continuum model of the simple dielectric fluid: Consistency between density based and continuum mechanics  methods}

\author{M. Sprik \thanks{Email: ms284@cam.ac.uk}}
\affiliation{Department of Chemistry, University of Cambridge,
  Lensfield Road, Cambridge CB2 1EW, United Kingdom}

\begin{abstract}
The basic continuum model for polar fluids is deceptively simple. The free energy integral  consists of four terms: The coupling of polarization to an external field, the electrostatic energy of the induced electric field interacting with itself and the stored polarization energy quadratic in the polarization.  A local function of density accounts for the mechanical state of the fluid. Viewed as a non-equilibrium free energy functional of number density and polarization, minimization in these two densities under constraints of the Maxwell field equations should lead the correct equilibrium state. The alternative is a continuum mechanics approach in which the mechanical degree of freedom is extended to full deformation.   We show that the continuum electromechanics method leads to a force balance equation which is consistent with the density functional equilibrium equation.  The continuum mechanics procedure is significantly more demanding. The gain is  a well defined pressure tensor derived from deformation of total  energy. This resolves the issue of the uncertainty in the pressure tensor  obtained  from integration of the force density, which is the conventional method  in density based thermomechanics. Our derivation is based on the variational electrostatics approach developed by Ericksen (Arch. Rational Mech. Anal. {\bf 183} 299 (2007)).
\end{abstract}

\maketitle

\section{Introduction} \label{sec:intro}
Free energy functionals for polarization fluctuations were introduced by Marcus to account for non-equilibrium solvation driving electron transfer in polar solvents\cite{Marcus1956,Marcus1994}. Non-equilibrium polarization functionals in the more recent literature are often referred to as Marcus-Felderhof functionals, adding the name of Felderhof who carried out a systematic study of their properties and foundation in linear response theory\cite{Felderhof1977,Bagchi1991}. Non-equilibrium polarization continues to play an important role in the theory of reaction dynamics in polar solvents\cite{Hynes1988}(For  a recent review see Dinpajooh et al.~\cite{Matyushov2017}).  

The Marcus-Felderhof  functional also found application in equilibrium continuum  theory as it can be used for variational determination of polarization in a dielectric medium interacting with fixed charge distributions\cite{Matyushov2017,Borgis2001,Attard2003,Maggs2006}. Computation of equilibrium polarization in complex geometries indeed is a major challenge. The long range nature of the dipolar tensor makes direct minimization of polarization cumbersome and expensive\cite{Hansen2001}. Improving the efficiency and numerical stability of these calculations has become a focus of research relevant for a large variety of applications such as colloidal and interface science, electrochemistry and biophysics. One option popular in physical chemistry  is to turn the integration of the electrostatic interactions into a boundary value problem optimizing the polarization surface charges instead of the volume polarization\cite{Hansen2001,Eisenberg2004,Frisch2010,Singer2016}.    

While the representation of polarization in terms of surface and interface charges is mathematical valid and physical meaningful\cite{Matyushov2014}, an alternative is to exploit the power (and inspiring elegance) of  Maxwell theory and describe the long-range electrostatic interactions  in terms of fields satisfying local differential equations.  The idea would be to rewrite the non-local interaction between polarization at different locations as an integral over the square of the local electric field generated by the polarization (its longitudinal component\cite{Matyushov2014,Matyushov2017}). The electric field is determined by solving the Maxwell equations converting the non-local integral equation to a set of coupled local (partial) differential equations\cite{Hansen2001,Maggs2012}.  The field equation for the Maxwell electric field requiring it to be curl free is readily satisfied by representing it as the gradient of a potential. That leaves the Maxwell equation for the dielectric displacement field providing the  coupling to the polarization and external charge. This approach amounts to  a Maxwell field formulation of the Marcus-Felderhof functional.

Ideally one would like  to treat the electric potential as an additional variational degree of freedom.   Hopefully the corresponding Euler-Lagrange equation would be equivalent to the Maxwell equation for the dielectric field. This hope is frustrated by a curious complication. The proper field equation for dielectric displacement can only be recovered  from a Legendre transform of the energy functional, which, unfortunately, changes the minimum  to a saddle point which makes the procedure unsuitable as for application  in numerical schemes\cite{Hansen2001,Maggs2006}. A similar difficulty was encountered in the quest for a variational solution of the Poisson-Boltzmann equation for ionic solutions where the problem acquired a certain notoriety\cite{Maggs2004,Maggs2012}. An ingenious solution proposed by Anthony Maggs is to impose the Maxwell equations by means of constraints implemented by the method of undetermined Lagrange multipliers\cite{Maggs2006}. 

Variational electrostatics is pursued in many diverse disciplines sometimes with little overlap. A development parallel to the activity in the physical chemistry of fluids took place in the field of continuum mechanics of solids\cite{Ericksen2007,Liu2013}. Of particular interest is the approach of J.~L.~Ericksen who reconsidered the use of an extended variational scheme with the dielectric displacement as basis rather than the electric field\cite{Ericksen2007}. The divergence of dielectric displacement (also called electric induction) vanishes in a dielectric continuum without embedded external charge and can therefore be represented as the curl of a vector potential, just as the magnetic induction. Using a convenient form of the energy functional Ericksen found that under conditions of stationarity for variation of this vector potential the curl of the electric field is zero, as required by the conjugate Maxwell equation\cite{Ericksen2007}. Moreover, being  an expression of the original Marcus-Felderhof functional in different variables, the energy functional remains convex.   Even if somewhat exotic, the Ericksen scheme has definite advantages, certainly  in formal derivations (see e.g.~Ref.~\citenum{Trianta2018}). It is the method adopted here and will be explained in explicit mathematical detail in section \ref{sec:elestat}. 

Ericksen developed his method as part of the continuing effort of merging electrostatics and elasticity theory.  This research program stretching over more than 50 years was initiated by two seminal papers by Toupin\cite{Toupin1956,Toupin1960}(for recent reviews see for example Refs.~\citenum{Suo2008,Ogden2009,Ogden2017}). The theoretical challenge in continuum electromechanics goes well beyond the problems faced in variational electrostatics of rigid systems. Now, in addition to the two Maxwell equations, a third field equation must be satisfied, namely force balance (the Cauchy equation). The central quantity in continuum mechanics is not force density but the stress tensor.  The key question is therefore what is the appropriate stress tensor for electroelastic systems. As can be expected, this will be combination of a form of the Maxwell stress tensor and mechanical stress due to short range (contact) interactions. However this partition is not unique leading to different expression for total stress\cite{Ogden2009}. Toupin derived his electromechanical stress tensor applying the principle of virtual work (for a gentle introduction we recommend Ref.~\citenum{Suo2008}). Ericksen went back to this work trying to derive the Toupin stress using minimum energy methods\cite{Ericksen2007}. 

 Stress  in linear elasticity theory  is the work conjugate of strain. However, applications to soft matter,   including elastic fluids, are based on non-linear elasticity theory originally developed for materials such as rubber.  Elasticity theory for finite deformation is usually formulated in a Lagrangian framework\cite{Ogden1997}. This raises the question what is the Lagrangian (material) form of the electric field and dielectric displacement.  This was resolved by Dorfmann and Ogden\cite{Ogden2005}  building on earlier work on electro-acoustics by Lax and Nelson\cite{Lax1976}.   This is the methodology used in this paper. Lagrangian electromechanics is unfamiliar  for most physical chemists with a statistical mechanics background.  For this reason, after specifying the continuum model system in section \ref{sec:Fdiel}, we present in section \ref{sec:chemech} our main result in the language of the classical density functional theory (DFT) of liquids.   The following technical sections (\ref{sec:Lagrange}-\ref{sec:varF}) fill in the rather heavy mathematical detail. We conclude in  section \ref{sec:disc} with a discussion.  

Summarizing the aim of this work, the question we set out to answer is whether the  differential equation for equilibrium density is consistent with the continuum mechanical equation for deformation  for the model simple dielectric fluid considered. The  reason that this might be in doubt is that the energy of polarized dielectric systems is sensitive to volume conserving (shear) deformations which are not accessed in an approach restricted to coupled density-polarization variation.  A lenghty derivation shows that this concern is not justified. The equilibrium density and polarization solving the Euler Langrange equation for these variables are also solutions  of the force balance equation of continuum mechanics (the electromechanical Cauchy equation). The center piece in this argument  is a systematic derivation of the dielectric stress tensor using the Langrangian electromechanical theory of Ref.~\citenum{Ogden2005} applied in the variational framework developed by Ericksen\cite{Ericksen2007}.   This yields an expression for the stresss tensor in agreement with the one obtained by Ericksen using his own slightly different methodology. Bringing this  particular dielectric  stress tensor and its derivation to the attention of the physical chemistry  community is the more practical aim of this investigation.

\section{System definition and statement of main result} \label{sec:system}

\subsection{Free energy of the simple dielectric fluid} \label{sec:Fdiel}
 Taking density $\rho$ and polarization $\mathbf{p}$ as the primitive variables  the continuum model of the simple dielectric fluid is specified  by  a functional  
\begin{equation}
 \mathcal{F}_{\textrm{d}}[\rho, \mathbf{p}] = \mathcal{F}_M[\rho] +
 \mathcal{F}_P[\rho, \mathbf{p}] + \mathcal{E}_F[\mathbf{p}]
 \label{eqn:Fdiel}
\end{equation}
$\mathcal{F}_M$  defines  the purely mechanical component of $\mathcal{F}_d$.   
\begin{equation}
\mathcal{F}_M[\rho] = \mathcal{F}_{sr}[\rho] + 
\int_{\Omega} v_{ext}(\mathbf{r}) \rho(\mathbf{r}) dv
\label{eqn:FM}
\end{equation}
 $\mathcal{F}_{sr} $ describes the short range interactions. In a ``simple'' fluid  it is the regular integral over a local free energy density $\phi(\rho)$
 \begin{equation} 
\mathcal{F}_{sr}[\rho] = \int_{\Omega} \phi(\rho) dv 
 \label{eqn:Fsr}
\end{equation}
where $\Omega$ is the volume of the body of liquid ($dv = d^3 \mathbf{r}$ is an infinitesimal volume element).  The second term in Eq.~\ref{eqn:FM} is the usual coupling to an external potential $v_{ext}$.
Assuming linear dielectrics, $\mathcal{F}_{P}$ is the stored polarization energy written as
\begin{equation} 
\mathcal{F}_{P}[\rho, \mathbf{p}] = \int_{\Omega} \frac{\mathbf{p}^2}{2 \chi(\rho)} dv 
 \label{eqn:Fp}
\end{equation}
The susceptibility $\chi(\rho)$ is allowed to vary with density (electrostriction). 

The last term in Eq.~\ref{eqn:Fdiel} is the electrostatic energy $\mathcal{E}_F$.  In the formulation of Refs.~\citenum{Felderhof1977} and \citenum{Attard2003} the electrostatic energy consists of the coupling of the polarization to the external field and an integral of the dipolar tensor coupling polarization at different locations. This form of $\mathcal{E}_F$ stays close to particle based statistical mechanics and the density functional theory (DFT) of polar liquids. Ericksen, however, starts from the total electrostatic of energy  Maxwell-Lorentz field theory\cite{Kovetz2000}
\begin{equation}
\mathcal{E}_F = \int_V \frac{\epsilon_0\mathbf{e}^2}{2} dv
\label{eqn:EML}
\end{equation}
 where $\mathbf{e}$ is the Maxwell electric field and $\epsilon_0$ the dielectric permittivity of vacuum.
 The energy $\mathcal{E}_F$ includes the electrostatic energy of the vacuum field. Separating this energy out we can write
\begin{equation} 
\mathcal{E}_F = \int_V \frac{\epsilon_0\mathbf{e}_0^2}{2} dv + \mathcal{U}_F
\label{eqn:UML}
\end{equation}
$\mathcal{U}_F$ plays the role of the system field energy which should vanish when the dielectric material is removed. The expression Ericksen uses for  $\mathcal{U}_F$  is an integral over an energy
density $e_{\textrm{E}}$
\begin{equation}
\mathcal{U}_F = \int_{V} e_{\textrm{E}}(\mathbf{p}) dv \equiv \mathcal{E}_{\textrm{E}} 
 \label{eqn:Ediel}
\end{equation}
with $e_{\textrm{E}}$ given by\cite{Ericksen2007}
\begin{equation}
  e_{\textrm{E}}(\mathbf{p}) = -\mathbf{p} \cdot \mathbf{e}_0 + \frac{\epsilon_0\hat{\mathbf{e}}^2 }{2}
  \label{eqn:Ekel} 
\end{equation}
where  $\hat{\mathbf{e}}$ is response field.  $\hat{\mathbf{e}}$ is related to the Maxwell field $\mathbf{e}$ and  applied field $\mathbf{e}_0$ as
\begin{equation}
  \mathbf{e} = \mathbf{e}_0 + \hat{\mathbf{e}}
  \label{eqn:eMax}
\end{equation}
$\hat{\mathbf{e}}$ is commonly referred to as the self field (We have adopted the compact hat notation of Ref.~\citenum{Trianta2018} to distinguish  between self field and Maxwell field).

 The internal electrostatic field energy density of  Eq.~\ref{eqn:Ekel} is specific for finite body geometry.  The dielectric occupying a  finite volume  $\Omega$ is enclosed in a container of total volume $V \gg \Omega $ (vacuum when empty).  While polarization is confined to the body ($\mathbf{p}=0$ outside) the self field $\hat{\mathbf{e}}$ spills out and is  nonzero in a region surrounding the body.  That is why the integration in Eq.~\ref{eqn:Ediel} is extended over the full volume $V$ of the container. A further condition is that the body is a pure dielectric  without embedded external charge. The polarizing external field $\mathbf{e}_0$ is a vacuum field. As a result the dielectric displacement $\mathbf{d} = \epsilon_0 \mathbf{e} + \mathbf{p}  $ is entirely transverse
\begin{equation}
         \ddiv \mathbf{d} = 0
  \label{eqn:divd}
\end{equation} 
  Eq.~\ref{eqn:divd} is complemented by the  Maxwell equation for the electric field
  \begin{equation}
         \curl \mathbf{e} = 0
  \label{eqn:curle}
\end{equation} 
which is generally valid.  Energy densities of the form of Eq.~\ref{eqn:Ekel} have been used by other authors\cite{Matyushov2017}. Ericksen however gives an original  derivation of this expression strictly based on Maxwell-Lorentz continuum theory\cite{Kovetz2000} without appealing to dipole densities\cite{Ericksen2007}. We will return to this important issue in section \ref{sec:elestat}.

Finally a comment on state variables.  The self field $\hat{\mathbf{e}}$ in Eq.~\ref{eqn:Ekel} is not an independent variable but an implicit functional  $\hat{\mathbf{e}}[\mathbf{p}]$ of polarization. Making this dependence explicit, $\hat{\mathbf{e}}$  is the longitudinal component of $\mathbf{p}$ as can in principle be determined by applying the Helmholtz vector field decomposition theorem(see e.g.~Refs.~\cite{Matyushov2017,Matyushov2014}).  As such, the self field $\hat{\mathbf{e}}$ is invariant under changes in the local density $\rho$ at fixed $\mathbf{p}$. However, it crucially  depends on the boundary (shape) of a dielectric body (even though we are dealing with a liquid  the shape is assumed to be fixed). Indeed the self field of a uniformly polarized body ($\ddiv \mathbf{p}=0$) is entirely determined by the surface term in the Helmholtz theorem.

\subsection{Chemical versus mechanical equilibrium} \label{sec:chemech}
A variational solution of the continuum model specified  in section \ref{sec:Fdiel} requires minimization of the free energy density Eq.~\ref{eqn:Fdiel} in $\rho$ and $\mathbf{p}$  under the constraint of the Maxwell field equations Eq.~\ref{eqn:divd} and \ref{eqn:curle}. As the applied field $\mathbf{e}_0$ is a vacuum field these conditions are transferred to the self fields 
\begin{eqnarray}
  \curl \hat{\mathbf{e}} &  = &  0  
\label{eqn:Maxe} \\
\ddiv \hat{\mathbf{d}}  & = & 0 
 \label{eqn:Maxd}
\end{eqnarray}
with the Maxwell-Lorentz equation \begin{equation}
  \hat{\mathbf{d}}  =  \epsilon_0 \hat{\mathbf{e}} + \mathbf{p}
  \label{eqn:hatLorentz} 
\end{equation}
relating $\hat{\mathbf{e}}$ and $\hat{\mathbf{d}}$ to the polarization $\mathbf{p}$. In addition appropriate jump conditions at the dielectric-vacuum boundary will have to be taken into account. 

As indicated in section \ref{sec:intro} density variation coupled to Maxwell field equations is a hard problem. The way this was solved by Erickson will be explained in section \ref{sec:elestat}. The result is as expected.  The Euler-Lagrange equation for the density can be expressed in a form familiar from DFT\cite{Evans1979}. Lumping the local energy densities in Eqs.~\ref{eqn:Fsr} and \ref{eqn:Fp} together in a single free energy density 
 \begin{equation}
  \psi(\rho, \mathbf{p}) = \phi(\rho) + \frac{\mathbf{p}^2}{2 \chi(\rho)}
  \label{eqn:psimple}
\end{equation}
we have at equilibrium  
 \begin{equation}
 \left(\frac{\partial \psi}{\partial \rho}\right)_{\mathbf{p}} = \mu - v_{ext}
 \label{eqn:ELrho}
 \end{equation}
 where the constant $\mu$ is the chemical potential for exchange of molecules with a reservoir or the Lagrange multiplier for imposing a fixed number of molecules.  Note that the partial derivative in Eq.~\ref{eqn:ELrho} is to be carried out at constant polarization. The Euler-Lagrange equation for polarization is the constitutive relation of linear dielectrics
 \begin{equation}
   \mathbf{p} = \chi \mathbf{e}
   \label{eqn:pchie}
 \end{equation}
 where $\mathbf{e}$ is the Maxwell field of Eq.~\ref{eqn:eMax} containing contributions from both the external and self field. 
 
 Before addressing the question of mechanical equilibrium we note that Eq.~\ref{eqn:ELrho} can be interpreted  as a chemical equilibrium condition. In local thermodynamics  the chemical potential is related to the free energy density as 
 \begin{equation}
  \mu_{\psi} (\rho) = \frac{\partial \psi}{\partial \rho}
  \label{eqn:mupsi}
 \end{equation}
 and hence according to Eq.~\ref{eqn:ELrho} at equilibrium 
 \begin{equation}
  \mu_{\psi} (\rho) = \mu -v_{ext}
  \label{eqn:chemeq}
 \end{equation}
 The local chemical potential is equal to an intrinsic chemical potential $\mu-v_{ext}(\mathbf{r})$ imposed by the interaction with the environment.  Is the system also in mechanical equilibrium? To check this we express the local pressure in terms of the local chemical potential   
 \begin{equation}
 P_{\psi} = \rho \mu_{\psi} - \psi 
 \label{eqn:Ppsi}
 \end{equation}
  Applying the chain rule we find for the spatial gradient
 \begin{equation}
\grad  P_{\psi} = \rho  \,\grad \mu_{\psi} 
\label{eqn:fpsi}
 \end{equation}
 where we have used $\mu_{\psi} \, \grad \rho = (\partial \psi /\partial \rho ) \grad \rho  = \grad \psi$.  The $\grad \psi$ terms cancel.  At equilibrium Eq.~\ref{eqn:chemeq} sets  $\grad \mu_{\psi} $ to  $ -\grad v_{ext}  $ yielding the force balance equation
 \begin{equation}
\grad  P_{\psi} =  \mathbf{f}_{ext}
\label{eqn:fbpsi}
 \end{equation}
 with $\mathbf{f}_{ext}$ the force density due to the action of the external potential
 \begin{equation}
 \mathbf{f}_{ext} = - \rho \,\grad v_{ext}
 \label{eqn:fext}
 \end{equation}
  In continuum mechanics $\mathbf{f}_{ext}$ is referred to as a body force.  Eq.~\ref{eqn:fbpsi} is interpreted as a force balance between an internal force $- \grad  P_{\psi} $ and body force $\mathbf{f}_{ext}$.
  
  The internal force density  $ -\grad P_{\psi}$ corresponding to the free energy density $\psi$ of  Eq.~\ref{eqn:psimple}  is well-known in continuum mechanics of dielectric material.  Working out the gradient  using equation  Eq.~\ref{eqn:fpsi} we find 
 \begin{equation*}
-\grad P_{\psi}   = - \rho \,\grad \frac{ \partial \phi}{\partial \rho} + 
 \rho \,\grad \left( \frac{\mathbf{P}^2 }{2 \chi^2} \frac{\partial \chi}{ \partial \rho} \right)
  \end{equation*}  
 The first term is the expected  force density due to the short range interactions. The second term is a force density generated by dielectric  polarization. Substituting the constitutive relation Eq.~\ref{eqn:pchie} and applying the chain rule we obtain
   \begin{equation}
  -\grad P_{\psi}   = - \grad P_{sr} -\frac{\mathbf{e}^2}{2} \grad \chi +
   \grad\left(  \rho \left(\frac{\partial \chi}{ \partial \rho} \right)\frac{\mathbf{e}^2}{2}\right)
   \equiv  \mathbf{f}_{\textrm{KH}} 
 \label{eqn:fKH}
   \end{equation}
 where $P_{sr} $ is the local  mechanical pressure. We recognize  the Korteweg-Helmholtz force density $\mathbf{f}_{\textrm{KH}}$  derived in many texts\cite{Suo2008,Landau1984,Melcher1989}.
 
 What would be different  if the force balance is obtained not from variation in density but from variation in deformation? The reader will have noticed a somewhat counterintuitive feature of the electrostatic energy density $e_{\textrm{E}}$ of Eq.~\ref{eqn:Ekel}. It depends on polarization only, not on density. Phrased in DFT language  the density derivative of the free energy functional Eq.~\ref{eqn:Fdiel}  is missing a contribution from the electrostatic energy. As a result
 \begin{equation}
 \frac{ \delta \mathcal{F}_d}{\delta \rho} = \psi[\rho,\mathbf{p}]
 \label{eqn:dFdrho}
 \end{equation}  
 where $\psi$ is the local free energy density of Eq.~\ref{eqn:psimple} leading to the seemingly truncated Euler-Lagrange Eq.~\ref{eqn:ELrho} for the density. The interaction with the electric field is accounted for indirectly through the constitutive relation Eq.~\ref{eqn:pchie}. This is what is remedied in continuum mechanics. The electrostatic energy is sensitive to  deformation and the corresponding mechanical Euler-Lagrange equation does contain a  contribution from the electrostatic field energy.  
 
 The derivation is lengthy and complicated filling the many pages of the following sections. The key step is determination of the stress tensor $\boldsymbol{\sigma}_{d}$ associated with the functional Eq.~\ref{eqn:Fdiel}. 
 \begin{equation}
\boldsymbol{\sigma}_{d} = \hat{\boldsymbol{\sigma}}_{\mathrm{T}} 
 - \frac{1}{2} \left(\mathbf{p} \cdot \mathbf{e} \right) \mathbf{I} + \left( \frac{\rho}{2}
 \left(\frac{\partial \chi}{\partial \rho}\right)  \mathbf{e}^2   - P_{sr}\right) \mathbf{I} 
\label{eqn:sigmad}
\end{equation}
where $\hat{\boldsymbol{\sigma}}_{\mathrm{T}}$ is the Toupin dielectric stress tensor\cite{Toupin1956}
\begin{equation}
 \hat{\boldsymbol{\sigma}}_{\mathrm{T}}  = \hat{\mathbf{d}} \otimes \hat{\mathbf{e}}
  - \frac{\epsilon_0}{2} \left(\hat{\mathbf{e}} \cdot \hat{\mathbf{e}} \right) \mathbf{I}
\label{eqn:lhatstress}
 \end{equation}
$\hat{\boldsymbol{\sigma}}_{\mathrm{T}}$ has the familiar form of a stress tensor in a dielectric continuum\cite{Landau1984} formed however from the self fields $\hat{\mathbf{d}}$ and  $\hat{\mathbf{e}}$ rather than the corresponding total  fields $\mathbf{d}$ and  $\mathbf{e}$.  Eq.~\ref{eqn:sigmad} is the total stress tensor obtained by Ericksen in his 2007 paper\cite{Ericksen2007} using his electrostatic field energy density Eq.~\ref{eqn:Ekel}.  This  result will be reproduced in the following technical sections using the Lagrangian formalism of Ref.~\citenum{Ogden2005}.
 
 Equilibrium in continuum mechanics is formulated in terms of a force balance equation for volume (bulk) forces and surface forces (tractions).   The balance equation for the volume forces is the Cauchy equation, which for our simple dielectric fluid is written as 
 \begin{equation}
   \ddiv \boldsymbol{\sigma}_{\textrm{d}}  + \mathbf{f}_I  + \mathbf{f}_{ext} = 0
\label{eqn:dCauchy}
\end{equation}
$\boldsymbol{\sigma}_{\textrm{d}}$ is the stress tensor of Eq.~\ref{eqn:sigmad}.  $\mathbf{f}_I$ is the density of the Kelvin force exerted  by a non-uniform external field  acting as an electric body force. 
\begin{equation}
\mathbf{f}_I  = \left( \mathbf{p} \cdot \grad \right)\mathbf{e}_0 
\label{eqn:fI}
\end{equation}
 There is a second set of Cauchy equations for force couples  ensuring angular momentum conservation.  The second Cauchy equation is satisfied for symmetric stress tensors.  This symmetry requirement seems to be violated by the dyadic tensor product in the expression Eq.~\ref{eqn:lhatstress}.  Symmetry is however reinstated under conditions of the simple linear  constitutive relation Eq.~\ref{eqn:pchie} which was used to derive Eq.~\ref{eqn:sigmad}.
 
 Evaluating  the divergence of  $\boldsymbol{\sigma}_{\textrm{d}}$ (the details are given in the following technical sections) we find
 \begin{equation}
\ddiv \boldsymbol{\sigma}_{\textrm{d}} = \mathbf{f}_{\textrm{KH}}  - \mathbf{f}_I
\label{eqn:dforce}
\end{equation}
$\mathbf{f}_{\textrm{KH}}$ is the same  Korteweg-Helmholtz force density of Eq.~\ref{eqn:fKH}.
Substituting in Eq.~\ref{eqn:dCauchy} the Kelvin force cancels and we are left with $\mathbf{f}_{\textrm{KH}}  + \mathbf{f}_{ext}= 0$ which is the force equilibrium as imposed  by DFT. The equilibrium DFT density and polarization are a solution of the Cauchy equation.  Coupled variation of density and polarization is the preferred method in physical chemistry\cite{Onuki2004,Tsori2011,Berthoumieux2018}.  The conclusion of the present study is that, applied to the simple dielectric fluid energy functional Eq.~\ref{eqn:Fdiel},  the equilibrium density and  polarization obtained by this approach are consistent with a full electromechanical treatment of the same system. However, the often rather ad hoc stress tensor used in physical chemistry studies is not always consistent with the energy functional.  The correct form, we believe, is the modified Toupin stress tensor of Eqs.~\ref{eqn:sigmad} and \ref{eqn:lhatstress} which was derived using the more elaborate methods of continuum electromechanics. 
 
 \section{Variational electrostatics of rigid dielectric continua} \label{sec:elestat}
 \subsection{Vector potential as additional variational parameter}
 The displacements field in pure dielectric material is transversal (Eq.~\ref{eqn:divd}). The variational scheme developed by Ericksen exploits this special property focusing on the dielectric self displacement $\hat{\mathbf{d}}$  instead of the electric self field $\hat{\mathbf{e}}$.   First,  the field energy of Eq.~\ref{eqn:Ekel} is expressed in terms of $\hat{\mathbf{d}}$ and the polarization $\mathbf{p}$ using Eq.~\ref{eqn:hatLorentz}
\begin{equation}
 {e}_{\textrm{E}}(\mathbf{p})
  =   -\mathbf{p} \cdot \mathbf{e}_0 + \frac{1}{2\epsilon_0}
 \left(\hat{\mathbf{d}} - \mathbf{p}\right)^2
\label{eqn:Ekeld}
\end{equation}
The self displacement field is divergence free (Eq.~\ref{eqn:Maxd}). Ericksen imposes this constraint  by representing $\hat{\mathbf{d}}$ in terms of a vector potential $\hat{\mathbf{a}}$
\begin{equation}
  \hat{\mathbf{d}} = \curl \hat{\mathbf{a}}
  \label{eqn:dcurla}
\end{equation} 
The vector field $\hat{\mathbf{a}}$ is treated as a second independent  electric variational parameter in addition to $\mathbf{p}$. The corresponding two-variable electrostatic energy density $\tilde{e}_{\textrm{E}}(\mathbf{p}, \hat{\mathbf{a}})$ is found by substituting Eq.~\ref{eqn:dcurla} in Eq.~\ref{eqn:Ekeld}
\begin{equation}
 \tilde{e}_{\textrm{E}}(\mathbf{p}, \hat{\mathbf{a}}) = -\mathbf{p} \cdot \mathbf{e}_0 + 
 \frac{\left(\curl \hat{\mathbf{a}} - \mathbf{p}\right)^2 }{2\epsilon_0}
  \label{eqn:Ekela}
\end{equation}
with the corresponding extended electrostatic energy functional
\begin{equation}
\tilde{\mathcal{E}}_{\textrm{E}}[\mathbf{p}, \hat{\mathbf{a}}]
 = \int_V \tilde{e}_{\textrm{E}}  (\mathbf{p}, \hat{\mathbf{a}})\, dv
 \label{eqn:Ekintela}
\end{equation}
The reason that extremization of Eq.~\ref{eqn:Ekintela}  leads to a valid solution for the electrostatics is that the Euler-Lagrange equation for $\hat{\mathbf{a}}$  is equivalent to Maxwell equation for the self field  (Eq.~\ref{eqn:Maxe})\cite{Ericksen2007}. Changing $\hat{\mathbf{a}}$ to $\hat{\mathbf{a}} + \delta \hat{\mathbf{a}}$ keeping $\mathbf{p}$ fixed yields a first order change in electrostatic energy
\begin{equation*}
\delta \tilde{\mathcal{E}}_{\textrm{E}} = 
 \frac{1}{\epsilon_0} \int_V \left( \curl \hat{\mathbf{a}} - \mathbf{p}\right)  \cdot \curl \delta \hat{\mathbf{a}} \, dv
\end{equation*}
where we have used that $ \delta \left(\curl \hat{\mathbf{a}} \right) = \curl \delta \hat{\mathbf{a}}$.  Converting the integrand  using the vector identity Eq.~\ref{eqn:divuoutv} we can then apply the divergence theorem and obtain
\begin{equation}
\delta \tilde{\mathcal{E}}_{\textrm{E}} = \frac{1}{\epsilon_0} \int_V  \curl \left( \curl\hat{\mathbf{a}} - \mathbf{p}\right)  \cdot \delta \hat{\mathbf{a}} \, dv \\
   + \frac{1}{\epsilon_0} \int_{\partial V}  \mathbf{n}_{\partial V} \wedge \left( \curl \hat{\mathbf{a}} - \mathbf{p}\right)  \cdot \delta \hat{\mathbf{a}} \, ds
   \label{eqn:dwall}
\end{equation}
where $\partial V$ is the boundary of $V$ with normal $\mathbf{n}_{\partial V}$. While $\mathbf{p}$ is strictly zero beyond the periphery of a finite dielectric body  the vector potential $\hat{\mathbf{a}}$, similar to $\hat{\mathbf{e}}$ is not. However,  we can assume that  $\hat{\mathbf{a}}$ decays to zero with increasing distance from the dielectric body and can be neglected at the vacuum boundary $\partial V$ leaving only the spatial integral over  $V$. This assumption is justified if the dielectric carries no net charge (polarization charge ingrates to zero). Then with the usual argument in variational theory the integral can only vanish for arbitrary $\delta \hat{\mathbf{a}}$ if 
\begin{equation}
\curl \left(\curl\hat{\mathbf{a}} - \mathbf{p}\right) = 0
\label{eqn:curl2a}
\end{equation} 
validating  the identification 
\begin{equation}
 \hat{\mathbf{e}} = \left( \curl \hat{\mathbf{a}} - \mathbf{p} \right)/\epsilon_0
\label{eqn:eselfa}
\end{equation}
 with $\hat{\mathbf{e}} $ satisfying Eq.~\ref{eqn:Maxe}.

 As pointed out by Ericksen, there seems to be more to Eq.~\ref{eqn:curl2a} than this. Expanding the repeated curl using Eq.~\ref{eqn:curl2} we can write
\begin{equation}
\grad \left( \ddiv \hat{\mathbf{a}}  \right)
- \Delta \hat{\mathbf{a}} = \curl \mathbf{p}
 \label{eqn:helma} 
\end{equation}
 Eq.~\ref{eqn:helma} suggests to introduce a transverse gauge $\ddiv \hat{\mathbf{a}} = 0$ familiar from Maxwell theory for electromagnetic fields. The question is however whether this is compatible with Maxwell jump conditions at boundaries. Ericksen argues that this is indeed the case simplifying Eq.~\ref{eqn:helma} to
\begin{equation}
\Delta \hat{\mathbf{a}} =- \curl \mathbf{p}
 \label{eqn:laplaca}
 \end{equation}
 The vector potential $\hat{\mathbf{a}}$  can therefore be used as a mathematically convenient representation of transverse polarization. In special geometries where polarization is longitudinal $\hat{\mathbf{a}}$ is  a solution of the vacuum Poisson equation. 
 
 \subsection{Varying the polarization at fixed vector potential}

 In order to apply the dielectric vector potential scheme to the simple dielectric liquid its free energy functional of Eq.~\ref{eqn:Fdiel} is extended to 
 \begin{equation} 
  \tilde{\mathcal{F}}_{\textrm{d}}[\rho, \mathbf{p}, \hat{\mathbf{a}}] = 
\mathcal{F}_{sr}[\rho] + \mathcal{F}_P[\rho, \mathbf{p}] +
 \tilde{\mathcal{E}}_{\textrm{E}}[\mathbf{p}, \hat{\mathbf{a}}]
  \label{eqn:Fdiela}
\end{equation}
 with the electrostatic energy as given in Eq.~\ref{eqn:Ekintela}. The short range and polarization terms are not modified and are still equal to the integrals of Eq.~\ref{eqn:Fsr} respectively Eq.~\ref{eqn:Fp}. The crucial difference with the variational problem defined by Eq.~\ref{eqn:Fdiel} is that the variation in $\mathbf{p}$ is now carried out at fixed $\hat{\mathbf{a}}$.  This greatly simplifies the expression for the first order change of the electrostatic energy induced by a change of  polarization $\mathbf{p} \rightarrow \mathbf{p} + \delta \mathbf{p}$ which is
\begin{equation*}
\delta \tilde{\mathcal{E}}_{\textrm{E}}  = \int_{\Omega} \left(  - \mathbf{e}_0 -\frac{1}{\epsilon_0} \left(\curl \hat{\mathbf{a}} - \mathbf{p} \right) \right) \cdot \delta \mathbf{p} \, dv
\end{equation*}
Integration can again be limited to $\Omega$ because the polarization remains confined to the body.  To this we must add the first order change in $\mathcal{F}_P$ which leads to the Euler Lagrange equation
\begin{equation}
 \frac{\partial \psi}{ \partial \mathbf{p}}  = \mathbf{e}_0 +\frac{1}{\epsilon_0} \left(\curl \hat{\mathbf{a}}  -\mathbf{p} \right) 
 \label{eqn:pconst}
\end{equation}
where $\psi$ is the local energy density defined in Eq.~\ref{eqn:psimple}.
Substituting Eq.~\ref{eqn:eselfa} and combining the applied and self field field using Eq.~\ref{eqn:eMax}
 we recover the constitutive equation for polarization Eq.~\ref{eqn:pchie}.  Variation in density would give a third Euler-Lagrange. Because density is not appearing as an explicit variable in the electrostatic energy $\tilde{e}_{\textrm{E}}(\mathbf{p},\hat{\mathbf{a}}) $ we end up with the regular DFT equation  Eq.~\ref{eqn:ELrho}.  
 
The Euler-Lagrange equations of the Ericksen scheme are the dielectric equations we were aiming for. Legendre transformation of the dielectric energy functional Eq.~\ref{eqn:Fdiel} as required by schemes based on using the electric potential can therefore be avoided (see section \ref{sec:intro}). Ericksen's proof  that  the corresponding stationary state is also a minimum relies on Eq.~\ref{eqn:UML} with the internal energy $\mathcal{U}_F$ given by Eq.~\ref{eqn:Ediel} and energy density Eq.~\ref{eqn:Ekel}. The full Maxwell Lorentz energy Eq.~\ref{eqn:EML} is manifestly convex. The energy of the vacuum field is a constant and, hence the field energy functional $\mathcal{E}_{\textrm{E}}$ is convex. 
This desirable property is not immediately  obvious from Eq.~\ref{eqn:Ekel} and is also questionable for the extended dielectric  models (see e.g..~ Ref.~\citenum{Madden1984}) favoured in the computationally oriented literature in physical chemistry\cite{Felderhof1977,Borgis2001,Attard2003,Maggs2006,Matyushov2017}.   Let us therefore reiterate the conditions for the validity mentioned in section \ref{sec:Fdiel}. These equations apply for a finite dielectric body in a container. The container must be  large enough for the  self dielectric displacement field $\hat{\mathbf{d}}$ to decay sufficiently fast outside the dielectric so that it can be neglected at the boundaries $\partial V$ of the container. We already made use of this property when discarding the surface term in Eq.~\ref{eqn:dwall}.  For the detailed proof we refer to Ericksen's paper\cite{Ericksen2007} (see also  the discussion in Ref.~\cite{Ogden2009}). 
 
 \section{Transformation to an arbitrary reference frame} \label{sec:Lagrange}
 \subsection{Maxwell equations in Lagrangian representation} \label{sec:lamax}
 
  In the Lagrangian or material formulation of continuum mechanics the state of a  system is described in terms of a deformation from a reference state also called material state. The deformed state is called the current or spatial state. This is how constitutive stress-strain relations are  specified in the theory of non-linear elasticity\cite{Ogden1997}.   In fluid mechanics this is not necessary.  This, however, does not prohibit a Lagrangian description of equilibrium fluids  and, in fact, offers a way of deriving expressions for stress tensors using the established mathematical machinery of the theory of nonlinear elasticity (For an instructive example see Ref.~\citenum{Eremeyev2014}).  Obviously, the reference frame is now arbitrary  and any result, once transformed back to current space,  should have lost all traces of the reference frame.  
  
  First we list the necessary kinematics.  The central mathematical construct is a  vector function $\boldsymbol{\chi}(\mathbf{X}) $ mapping points $\mathbf{X}$ in the reference space  $B_r$ to points $\mathbf{x}$ in the current  space  $B$ so that $\mathbf{x} = \boldsymbol{\chi}(\mathbf{X})$.  Strain in  elastic solids is not directly equal to displacement  $\boldsymbol{\chi}(\mathbf{X}) $ but is computed from the deformation gradient tensor 
  \begin{equation}
   F_{i\alpha} = \frac{\partial x_i}{\partial X_{\alpha}}
   \label{eqn:Fcart}
 \end{equation}
 where Cartesian coordinates in current and  reference  space are written as  $(x_i, i=1,2,3)$ and   $(X_{\alpha}, \alpha =1,2,3)$, respectively.  In continuum mechanical theory direct vector notation is often more convenient. To change the notation of a derivative operator in current space ($\grad,\ddiv,\curl$)  to the corresponding derivative operator in reference space  the first letter of the operator name is switched to upper case. In this convention the deformation gradient tensor of Eq.~\ref{eqn:Fcart} is written as 
  \begin{equation}
 \mathbf{F} = \Grad \boldsymbol{\chi} =  \left( \nabla_X \mathbf{x}\right)^{\textrm{T}}
 \label{eqn:Fvect}
 \end{equation}
 where $\nabla_X$ is the nabla operator in the reference configuration and $^\mathrm{T}$ denotes the transpose (Note the difference between the $\Grad$ and $\nabla_X$ operators when acting on vectors. The order of the ($i,\alpha$) indices is  exchanged \cite{Eremeyev2014}). Gradients of $\mathbf{F}$ are converted according to
 \begin{equation}
 \nabla_X = \mathbf{F}^{\mathrm{T}} \nabla, \qquad 
 \nabla = \mathbf{F}^{\mathrm{-T}} \nabla_X
 \label{eqn:Ggrad}
 \end{equation} 
  with $\mathbf{F}^{\mathrm{-T}} = \left( \mathbf{F}^{\mathrm{-1}} \right)^{T}$.  Number density (in non-reactive systems) is conserved and hence the spatial density $\rho$ is related to the density $\rho_0$ in  reference space as
\begin{equation}
   J \rho = \rho_0
  \label{eqn:rho0}
 \end{equation}
 where $J$ is the Jacobian of the deformation gradient tensor
 \begin{equation}
    J = \det \mathbf{F} 
\label{eqn:JdetF}
 \end{equation}
 $J$ is positive.
 
  The electric field $\mathbf{e}$ and displacement field $\mathbf{d}$  in Maxwell theory are tied to their source charge densities by differential equations.  Position derivatives change switching over from the current to the reference frame (Eq.~\ref{eqn:Ggrad}) suggesting  that $\mathbf{e}$  and $\mathbf{d}$   cannot simply remain the same when expressed in the reference frame.   The central idea in Lagrangian electromechanics is to adjust $\mathbf{e}$  and $\mathbf{d}$ making the Maxwell equations form invariant\cite{Ogden2005}. The Lagrangian counterparts of the electric field $\mathbf{e}$ and displacement $\mathbf{d}$ are denoted by capital $\mathbf{E}$ and $\mathbf{D}$ respectively and are obtained from the fields in current space according to
 \begin{eqnarray}
   \mathbf{E} & = & \mathbf{F}^{\mathrm{T}} \mathbf{e}
\label{eqn:RE} \\
   \mathbf{D} & = & J \mathbf{F}^{-1} \mathbf{d}
\label{eqn:RD}
 \end{eqnarray}
With these transformation rules the dielectric Maxwell equations are written in the reference frame as  
 \begin{equation}
   \Curl \mathbf{E} = 0, \qquad \Ddiv \mathbf{D} = 0
\label{eqn:RMaxwell}
 \end{equation}
 where $\Curl$ and $\Ddiv$ are the curl and divergence operators applied in reference coordinates $\mathbf{X}$. The equations for the material electric and displacement field have the same form as in the current  frame.  Note also that $\mathbf{D} \cdot \mathbf{E} = J \mathbf{d} \cdot \mathbf{e}$. The inproduct of $\mathbf{d}$ and $\mathbf{e}$ is transformed as a density.  However,  non-equilibrium electrostatic energy is determined,  not  by $\mathbf{d} \cdot \mathbf{e}  $, but by $\mathbf{e} \cdot \mathbf{e}$, which is not behaving as a density under transformation. This is how in Lagrangian theory the geometry dependence of the energy is captured.
 
While the transformation rules for electric field and dielectric displacement are imposed by requiring the preservation (form invariance) of the Maxwell equations  the Lagrangian form $\mathbf{P}$ of polarization is open to some choice. We follow Dorfmann and Ogden\cite{Ogden2005,Ogden2009} and define $\mathbf{P}$ as   
 \begin{equation}
 \mathbf{P} = J \mathbf{F}^{-1} \mathbf{p}
\label{eqn:RP}
 \end{equation}
which is the same rule as for the displacement field (Eq.~\ref{eqn:RD}). Combining Eqs.~\ref{eqn:RE}, \ref{eqn:RD} and \ref{eqn:RP} gives
 \begin{equation}
 \mathbf{D} =  \epsilon_0 J \mathbf{C}^{-1} \mathbf{E} + \mathbf{P} 
\label{eqn:RLorentz} 
\end{equation}
 with $\mathbf{C}$ the right Cauchy-Green strain tensor. 
 \begin{equation}
    \mathbf{C} = \mathbf{F}^{\mathrm{T}} \mathbf{F}, \qquad 
    \label{eqn:CGstrain}
 \end{equation}
 We must accept that the fundamental Maxwell-Lorentz relation  $\mathbf{d} = \epsilon_0 \mathbf{e} + \mathbf{p}$ cannot be carried over from the current frame to the reference frame. This is inevitable because $\mathbf{e}$ and $\mathbf{d}$ transform according to different rules (Eqs.~\ref{eqn:RE} and \ref{eqn:RD}).  In this respect the relation $\mathbf{d} = \epsilon_0 \mathbf{e} + \mathbf{p}$, although universal,  is similar to a constitutive equation\cite{Kovetz2000}.    
 
 \subsection{Material specification of the simple dielectric fluid} \label{sec:LFdiel}
 The electric fields used in the variational treatment of the simple dielectric fluid in section \ref{sec:elestat} were not $\mathbf{e}$ and $\mathbf{d}$ but the self fields $\hat{\mathbf{e}}$ and $\hat{\mathbf{d}}$. Of course  the same transformation rules apply to the Lagrangian versions  of self fields. Indicating the Lagrangian self fields again in capitals keeping the hat we have
  \begin{eqnarray}
\hat{\mathbf{E}}& = & \mathbf{F}^{\mathrm{T}} \hat{\mathbf{e}} 
\label{eqn:RhatE} \\
\hat{\mathbf{D}} & = & J \mathbf{F}^{-1} \hat{\mathbf{d}}
\label{eqn:RhatD}
\end{eqnarray}
satisfying Maxwell equations in agreement with  Eq.~\ref{eqn:RMaxwell}
 \begin{equation}
  \Curl \hat{\mathbf{E}}  =  0, \qquad
  \Ddiv \hat{\mathbf{D}}  = 0  
\label{eqn:RhatMaxwell}
\end{equation} 
with the self field Lagrangian Lorentz equation Eq.~\ref{eqn:RLorentz}
\begin{equation}
\epsilon_0 \hat{\mathbf{E}} = J^{-1}  \mathbf{C} \left(\hat{\mathbf{D}} 
- \mathbf{P} \right)
\label{eqn:RhatLorentz}
\end{equation}
arranged in the form it will be applied in the following.

The task we have now is to express the free energy functional  as used in section \ref{sec:elestat} in material form. We do this  term by term starting with the electrostatic field energy   of $ e_{\textrm{E}}(\mathbf{p})$ of Eq.~\ref{eqn:Ekeld}. There are two contributions, a term coupling  polarization to the external field and the electrostatic self energy. It will be convenient to treat these terms separately. The interaction with the external field will be from now on indicated by  $ \mathcal{E}_{I}$  
\begin{equation}
 \mathcal{E}_I = - \int_V  \mathbf{p} \cdot \mathbf{e}_0 dv 
 \label{eqn:eI}
 \end{equation}
 Applying the inverse of Eqs.~\ref{eqn:RE} and \ref{eqn:RP} gives the material representation  of $\mathcal{E}_I $   which is a functional of the polarization $\mathbf{P}$ in the reference frame and deformation gradient $\mathbf{F}$
 \begin{equation}
 \tilde{\mathcal{E}}_I  \left( \mathbf{P}, \mathbf{F} \right) =  - \int_V J^{-1} 
 \left( \mathbf{F} \mathbf{P}  \right) \cdot \left( \mathbf{F}^{-\mathrm{T}} \mathbf{E}_0 \right)J  dV 
  =  - \int_V \mathbf{P} \cdot \mathbf{E}_0 \, dV
  \label{eqn:teI}
\end{equation}
$dV = J^{-1}dv$ is a reference space volume element . The factors $J$ cancel.  Similar to $\mathbf{d} \cdot \mathbf{e}$ the inproduct $\mathbf{p} \cdot \mathbf{e}$ also transforms as a density. Note, however, that $\mathbf{E}_0 = \mathbf{F}^{\textrm{T}}\mathbf{e}_0$ varies with deformation because $\mathbf{e}_0$ is fixed implying that  also the external field interaction Eq.~\ref{eqn:teI} is sensitive to changes in shape and will therefore contribute to the stress. 

Next we rewrite the self interaction energy for which we also introduce a separate symbol
\begin{equation}
\mathcal{E}_S = \int_V \frac{1}{2\epsilon_0}  \left(\hat{\mathbf{d}} - \mathbf{p}\right)^2 dv
\label{eqn:eself}
\end{equation}
Applying Eqs.~\ref{eqn:RE}, \ref{eqn:RD} and \ref{eqn:RP} we can write $\mathcal{E}_S$ as
\begin{eqnarray}
 \mathcal{E}_S & = & \int_v \frac{1}{2\epsilon_0} \left( J^{-1} \mathbf{F} 
 \left(\hat{\mathbf{D}} - \mathbf{P} \right) \right) \cdot
 \left( J^{-1} \mathbf{F} \left(\hat{\mathbf{D}} - \mathbf{P} \right) \right)
J dV \nonumber \\ & = & \int_V
 \frac{J^{-1}}{2\epsilon_0} \left( \hat{\mathbf{D}} - \mathbf{P} \right) \cdot
 \mathbf{C} \left(\hat{\mathbf{D}} - \mathbf{P} \right) dV 
\label{eqn:dpDP} 
\end{eqnarray}
where in the second step we have substituted Eq.~\ref{eqn:CGstrain} 
 In preparation for a repeat of the derivation of section \ref{sec:elestat} we represent $\hat{\mathbf{D}}$ in terms of a vector potential $\hat{\mathbf{A}}$. The field transformation rules Eqs.~\ref{eqn:RE} and \ref{eqn:RD} ensure that a solution to the Maxwell equations Eqs.~\ref{eqn:RhatMaxwell} also satisfy the original Maxwell equations in the current frame.  Therefore, in direct correspondence to Eq.~\ref{eqn:dcurla}, we set
\begin{equation}
  \hat{\mathbf{D}} = \Curl \hat{\mathbf{A}}
  \label{eqn:DcurlA}
\end{equation} 
and extend the self energy functional to a three variable energy functional 
\begin{equation}
\tilde{\mathcal{E}}_S [\mathbf{P}, \hat{\mathbf{A}}, \mathbf{F}]
= \int_V \frac{J^{-1}}{2\epsilon_0} \left( \Curl \hat{\mathbf{A}} - \mathbf{P} \right) \cdot
\mathbf{C}\left( \Curl \hat{\mathbf{A}} - \mathbf{P} \right) dV
\label{eqn:tReS}
 \end{equation}
 The sum
 \begin{equation}
 \tilde{\mathcal{E}}_{\textrm{E}} [\mathbf{P}, \hat{\mathbf{A}}, \mathbf{F}]
  = \tilde{\mathcal{E}}_I [\mathbf{P}, \mathbf{F}] + 
  \tilde{\mathcal{E}}_S [\mathbf{P}, \hat{\mathbf{A}}, \mathbf{F}]
 \label{eqn:tReE}
 \end{equation}
 is the continuum mechanics adaptation of Eq.~\ref{eqn:Ekintela}. 

Transformation of the polarization energy $\mathcal{F}_P$ (Eq.~\ref{eqn:Fp}) proceeds along the same lines. Consistent with $\tilde{\mathcal{E}}_S$ of Eq.~\ref{eqn:tReS}   the dependence on polarization  is specified in terms of the material representation $\mathbf{P}$  (Eq.~\ref{eqn:RP}) which  plays role of  independent variational  degree of freedom. This introduces  again  the strain tensor $\mathbf{C}$ as an effective coupling matrix. A further effect of deformation is through the susceptibility which varies with density.  Using Eq.~\ref{eqn:rho0}  we write $\chi(\rho)= \chi \left( J^{-1}\rho_0 \right)$. The result is  a material polarization energy functional
\begin{equation}
\tilde{\mathcal{F}}_P[\mathbf{P}, \mathbf{F}]   = 
  \int_V  \frac{J^{-1}}{2\chi(J^{-1} \rho_0)} \mathbf{P} \cdot \mathbf{C} \mathbf{P} dV
\label{eqn:ReP}
\end{equation}
The transformation of the local mechanical energy Eq.~\ref{eqn:Fsr} is standard.
\begin{equation} 
  \tilde{\mathcal{F}}_{sr}[\mathbf{F}] = \int_{\Omega_r} \phi(J^{-1}\rho_0) J dV
\label{eqn:Resr}
\end{equation}

\section{Varying the electric degrees of freedom} \label{sec:varAP}
We will now retrace the variational procedure of section \ref{sec:elestat} for the material electrical degrees of freedom at rigid geometry. Thus, changing $\hat{\mathbf{A}}$ to $\hat{\mathbf{A}} + \delta \hat{\mathbf{A}}$ keeping $\mathbf{P}$ and $\mathbf{F}$ fixed we determine the first order change in the electrostatic field energy Eq.~\ref{eqn:tReE}
\begin{equation*}
  \delta \tilde{\mathcal{E}}_{\textrm{E}} =  \delta \tilde{\mathcal{E}}_S  = \int_V 
 \frac{J^{-1}}{\epsilon_0} \mathbf{C}\left( \Curl \hat{\mathbf{A}} - \mathbf{P}\right) 
 \cdot \Curl \delta \hat{\mathbf{A}} \, dV
\end{equation*}
where have used that the Cauchy-Green strain tensor (Eq.~\ref{eqn:CGstrain}) is symmetric. Next applying the divergence theorem in reference space 
\begin{eqnarray*}
\delta \tilde{\mathcal{E}}_{\textrm{E}} & = & \int_V 
 \Curl \left( \frac{J^{-1}}{\epsilon_0} \mathbf{C} \left(
\Curl  \hat{\mathbf{A}} - \mathbf{P}\right) \right)
 \cdot \delta \hat{\mathbf{A}} \, dV
 \\ & & 
 +  \int_{\partial V}  \mathbf{N}_{\partial V} \times \left( \frac{J^{-1}}{\epsilon_0} 
\mathbf{C} \left( \Curl \hat{\mathbf{A}} - \mathbf{P}\right)  \right)
\cdot \delta \hat{\mathbf{A}} \, dS
\end{eqnarray*}
As in section \ref{sec:elestat} we assume that the self fields vanish at the container boundary $\partial V$ and we are left with the Euler-Lagrange equation
\begin{equation}
 \Curl \left( \frac{J^{-1}}{\epsilon_0} \mathbf{C} \left(
\Curl  \hat{\mathbf{A}} - \mathbf{P}\right) \right) = 0
\end{equation}
Substituting Eq.~\ref{eqn:DcurlA} and referring back to the modified Lorentz relation Eq.~\ref{eqn:RhatLorentz} verifies that the Euler Lagrange equation for variation in $\hat{\mathbf{A}}$ recovers the Maxwell equation $\Curl \hat{\mathbf{E}}= 0 $ for  the self field in the reference frame.  

Changing $\mathbf{P}$ to $\mathbf{P} + \delta \mathbf{P}$ keeping $\hat{\mathbf{A}}$ and $\mathbf{F}$ fixed both the interaction energy Eq.~\ref{eqn:teI} and self energy Eq.~\ref{eqn:dpDP} contribute to the first order variation of the field energy Eq.~\ref{eqn:tReE}
\begin{eqnarray*}
\delta \tilde{\mathcal{E}}_{\textrm{E}}  & = &  - \int_V \left(  \mathbf{E}_0 +
 \frac{J^{-1}}{\epsilon_0} \mathbf{C}\left( \Curl \hat{\mathbf{A}} - \mathbf{P}\right) 
 \right) \cdot  \delta \mathbf{P} \, dV
 \\ & = & 
 - \int_V \left(  \mathbf{E}_0 + \hat{\mathbf{E}}
 \right) \cdot  \delta \mathbf{P} \, dV
\end{eqnarray*}
where in the second step we have restored the explicit dependence on the material electric self field using Eqs.~\ref{eqn:DcurlA} and \ref{eqn:RhatLorentz}. Adding the variation in the stored polarization energy (Eq.~\ref{eqn:ReP})
\begin{equation*}
\delta \tilde{\mathcal{F}}_P  =  \int_V  \left(
 \frac{J^{-1}}{\chi} \mathbf{C} \mathbf{P} \right)  \cdot  \delta \mathbf{P} \, dV
\end{equation*}
 leads to the Euler-Lagrange equation
\begin{equation*}
\frac{J^{-1}}{\chi} \mathbf{C} \mathbf{P} = \mathbf{E}_0 + \hat{\mathbf{E}}
\end{equation*}
Resolving the Cauchy-Green tensor $\mathbf{C}$ using Eq.~\ref{eqn:CGstrain} gives
\begin{equation}
 J^{-1}\mathbf{F} \mathbf{P} = \chi  \mathbf{F}^{-\mathrm{T}}  \mathbf{E}
\label{eqn:RELP}
\end{equation}
The left hand side matches the inverse of the transformation Eq.~\ref{eqn:RP} for polarization and the right hand side the inverse of the transformation Eq.~\ref{eqn:RE} for electric fields. Applying these transformations we see that Eq.~\ref{eqn:RELP} reduces to the sought after constitutive relation Eq.~\ref{eqn:pchie}  in the current frame. Note also that the material constitutive relation between $\mathbf{P}$ and $\mathbf{E}$ is still linear, but with a susceptibility coefficient varying with deformation. For finite deformation the geometry dependence is even non-linear and similar to Eq.~\ref{eqn:RhatLorentz}.

\section{Variational continuum mechanics of simple liquids} \label{sec:varsimp}

\subsection{Variation of deformation} \label{eqn:varchi}
So far so good. We have reproduced in a roundabout way what we had already established in section \ref{sec:elestat} in an Eulerian framework. The ``pull back'' to a Lagrangian frame is a device  for variation of the geometry which will give the expression for  the stress tensor.  In nonlinear continuum mechanics deformation  is quantified as a  change  $\boldsymbol{\chi} \rightarrow \boldsymbol{\chi} + \delta \boldsymbol{\chi}$ of the placement in current  space\cite{Ogden1997}.  Indicating the change $\delta \mathbf{x}$ of position by $\mathbf{u}$ we therefore have to evaluate the first order differences  induced by making the  substitution
\begin{equation}
  \mathbf{x}(\mathbf{X}) \rightarrow\mathbf{x}(\mathbf{X}) + \mathbf{u}(\mathbf{X})
  \label{eqn:xdv}
\end{equation}
The Lagrangian form of the energies in section \ref{sec:LFdiel} were all functions of the deformation gradient tensor $\mathbf{F}$ defined in Eq.~\ref{eqn:Fvect} and its inverse. The first order differential
is the gradient of the  change $\mathbf{u}$ in displacement evaluated in the reference frame 
\begin{equation}
\delta \mathbf{F}  = \Grad \mathbf{u} = \left( \nabla_X \mathbf{u} \right)^{\textrm{T}},
\qquad
\delta \mathbf{F}^{-T}   = - \mathbf{F}^{-\mathrm{T}}  \cdot \delta \mathbf{F}^{\mathrm{T}}
\cdot \mathbf{F}^{-\mathrm{T}}
\label{eqn:dFdefm}
\end{equation}
Another important geometric differential  is the variation $\delta J$  of the determinant of $\mathbf{F}$ (Eq.~\ref{eqn:JdetF}). To find $\delta J$ we apply the chain rule after first recalling Jacobi's formula
\begin{equation}
 \frac{\partial \det \mathbf{F}}{\partial \mathbf{F}} =  \left(\det \mathbf{F}\right)\mathbf{F}^{-1}
\end{equation}
This gives
\begin{equation}
\delta J = J \mathrm{Tr} \left( \mathbf{F}^{-1} \delta \mathbf{F} \right)
= J \mathbf{F}^{-\textrm{T}} : \delta \mathbf{F}
\label{eqn:dRJ} 
\end{equation}
where  $\mathbf{A} : \mathbf{B} = \Sigma_{ij} A_{ij} B_{ij}$  stands for a double contraction of matrices $\mathbf{A}$ and $\mathbf{B}$.  With Eq.~\ref{eqn:rho0} we have therefore for the variation in density 
\begin{equation}
\delta \rho  = \delta \left(J^{-1} \rho_0 \right) = - \frac{\rho_0}{J^2} \delta J =
- \rho \mathbf{F}^{-\textrm{T}} : \delta \mathbf{F}
\label{eqn:dRrho}
\end{equation}
Returning to  the Eulerian representation we can write 
\begin{equation}
\delta J   =  J \ddiv \mathbf{u}  , \qquad \delta \rho  = - \rho \ddiv \mathbf{u} 
\label{eqn:drho}
\end{equation}
which is rather more recognizable\cite{Eremeyev2014}. We will also need the first order change in the Cauchy-Green matrix Eq.~\ref{eqn:CGstrain}.
\begin{equation}
\delta \mathbf{C}  =  \delta \left( \mathbf{F}^{\mathrm{T}} \mathbf{F} \right)
  =  \left( \delta  \mathbf{F}^{\mathrm{T}} \right) \mathbf{F} +
         \mathbf{F}^{\mathrm{T}} \delta  \mathbf{F} 
\label{eqn:dC}
\end{equation}
Eqs.~\ref{eqn:xdv}-\ref{eqn:dC} and further extensions are the  subject of the first section of almost every paper in non-linear continuum mechanics under the heading of kinematics. This includes the publications cite here\cite{Ericksen2007,Ogden2005,Ogden2009,Ogden2017,Suo2008,Trianta2018,Liu2014}. A paper we found particularly instructive in respect, although not on electromechanics but gradient capillary stress, is  Ref.~\citenum{Eremeyev2014}. For the continuum mechanics of simple liquids the equations in this section are all what is needed. 

\subsection{True and nominal stress} \label{eqn:stress}
Stress tensors in finite deformation continuum mechanics are obtained as derivatives of energy densities in reference space.  Differentiation is carried out with respect to material variables  and the result is  then transformed back to the current frame to give the physical (Cauchy) stress\cite{Ogden1997}.  Implemented in a variational framework this requires determining the differential of the free energy functional Eq.~\ref{eqn:Fdiel} in response to deformation.   As an illustration and for further reference will go (in minimal detail) through the application to the mechanical free energy $\mathcal{F}_M $ of Eq.~\ref{eqn:FM}. The material form of the integral over  local energy density $\phi(\rho)$ was given in Eq.~\ref{eqn:Resr}.  The derivation of the corresponding stress and work balance is general and  valid for all local density functions $\phi$.  In recognition of this we replace  ``\emph{sr}'' suffix by $\phi$.  

Expanding the integral Eq.~\ref{eqn:Resr}, now called $\delta \mathcal{F}_{\phi}$,  in differentials of $\phi$ and $J$  we get
\begin{equation} 
  \delta \mathcal{F}_{\phi}  = \int_{\Omega_r}
 \left( \frac{\partial \phi}{\partial \rho} J\delta \rho + \phi \delta J \right)  dV
 \label{eqn:dResr}
\end{equation}
Substituting Eqs.~\ref{eqn:dRJ} and \ref{eqn:dRrho} gives
\begin{equation} 
  \delta \mathcal{F}_{\phi}  =  \int_{\Omega_r}
  J \left(  - \rho\frac{\partial \phi}{\partial \rho}  + \phi \right)  
 \mathbf{F}^{-\textrm{T}} : \delta \mathbf{F} dV
 \label{eqn:dFesr}
\end{equation}
Eq.~\ref{eqn:dFesr}  has the form of incremental internal work expressed in material quantities
\begin{equation}
  W_{int} = \int_{\Omega_r} \boldsymbol{\Sigma}  : \delta \mathbf{F} dV
  \label{eqn:Pwork}
\end{equation}
 Eq.~\ref{eqn:Pwork} is fundamental in non-linear continuum mechanics defining the Lagrangian   stress tensor  $\boldsymbol{\Sigma}$. This quantity is also  referred to as the nominal\cite{Ogden1997} or  first Piola-Kirchhoff stress tensor. We  abbreviate this to  simply  the Piola stress tensor in the following.   The Piola stress tensor for the local mechanics specified by $\phi$ is therefore
 \begin{equation}
 \boldsymbol{\Sigma}_{\phi} = J \left(  - \rho\frac{\partial \phi}{\partial \rho}  + \phi \right)  
 \mathbf{F}^{-\textrm{T}} 
 \label{eqn:Piolasr}
 \end{equation}
 
 The Piola stress tensor $\boldsymbol{\Sigma}$ and the true (Cauchy) stress tensor $\boldsymbol{\sigma}$ in current space are related by a ``pull back'' and the inverse ``push forward'' transformation\cite{Ogden1997}
\begin{equation}
 \boldsymbol{\Sigma} = J \boldsymbol{\sigma}\mathbf{F}^{\textrm{-T}}, \qquad 
  \boldsymbol{\sigma} = J^{-1} \boldsymbol{\Sigma}\mathbf{F}^{\textrm{T}}
  \label{eqn:Piola}
\end{equation}
Inserting in Eq.~\ref{eqn:Pwork}  and using Eqs.~\ref{eqn:dFdefm} yields
\begin{equation*} 
 W_{int}   =   \int_{\Omega_r}
 \left( \boldsymbol{\sigma} \mathbf{F}^{-\textrm{T}} \right) : \delta \mathbf{F} J dV
   =  \int_{\Omega_r} \sigma : \left( \mathbf{F}^{-\textrm{T}}  \nabla_X \mathbf{u} \right) J dV
  \end{equation*}
  Using Eq.~\ref{eqn:Ggrad} this can be reverted to internal  work in Eulerian representation 
\begin{equation}
W_{int}   =  \int_{\Omega} \boldsymbol{\sigma} : \nabla \mathbf{u} \, dv =
 \int_{\Omega} \left( \grad \mathbf{u} \right) : \boldsymbol{\sigma}  \, dv
 \label{eqn:Cwork}
 \end{equation}

\subsection{External work and force balance equation } \label{sec:varclas}
While the external potential is fixed, geometric deformation will nonetheless affect the integral from which which the interaction its computed.  The change of the  interaction energy can be regarded as external work
 \begin{equation}
  W_{ext} =  \delta \int_{\Omega} v_{ext} \rho  dv
  \label{eqn:Wext}
 \end{equation}
 $W_{ext}$ can  be equally evaluated with the help of Eq.~\ref{eqn:dResr} if we take $\phi = \rho v_{ext} $. This gives
 \begin{equation*}
 W_{ext}  =  \int_{\Omega_r } \left( v_{ext}   J \delta \rho + v_{ext} \rho \delta J 
  + \delta v_{ext} J \rho \right) dV
 \end{equation*}
 The extra third term is because of the explicit spatial dependence of $v_{ext}$.    The first two terms cancel each other because $J \delta \rho = -\rho \delta J $  (see  Eq.~\ref{eqn:dRrho}) and therefore
 \begin{equation}
 W_{ext}   = \int_{\Omega} \rho  \left( \grad v_{ext} \right) \cdot \mathbf{u} dv =
 - \int_{\Omega}  \mathbf{f}_{ext} \cdot \mathbf{u} dv
 \label{eqn:bwork}
 \end{equation}
 where in the last step we have substituted Eq.~\ref{eqn:fext}. While this is the anticipated result we have gone through the derivation in some detail to verify that $W_{ext}$ is indeed the work done by the system on the environment, not the other way around. This observation will be important when we encounter Kelvin forces in section \ref{sec:elemech}.
 
 Finally we call on the variational principle requiring the  differential of the total energy to vanish
 \begin{equation*}
 \delta \left( \mathcal{F}_{\phi} + \int_{\Omega} \rho v_{ext} dv \right ) 
   = W_{int} + W_{ext} = 0
   \label{eqn:varmech}
 \end{equation*} 
 This should hold for arbitrary variation in displacement  $\mathbf{u}$. The external work $W_{ext}$ of Eq.~\ref{eqn:bwork} is already an differential in $\mathbf{u}$, but the internal work $W_{int}$ of Eq.~\ref{eqn:Cwork} is not (yet). It is given this form by application of the divergence theorem
 \begin{equation*}
 W_{int} = 
 \int_{\partial \Omega} \left (\bold{\sigma} \mathbf{n} \right) \cdot \mathbf{u} \, da - 
  \int_{\Omega}  \left( \ddiv \boldsymbol{ \sigma} \right)  \cdot \mathbf{u} \,   dv
 \end{equation*}
 $\mathbf{n}$ is the outward surface normal at the boundary $\partial \Omega$ of volume $\Omega$. Since $\mathbf{u}$ is arbitrary stationarity should apply to  the volume and surface integral independently. For the volume integral this leads to the Cauchy equation
 \begin{equation}
   \ddiv \boldsymbol{\sigma} + \mathbf{f}_{ext} = 0 
   \label{eqn:Cauchy}
 \end{equation}
In mechanical equilibrium  the divergence of the ``true'' (Cauchy) stress must match minus the external force density. $\ddiv \boldsymbol{\sigma}$  can thus be interpreted as an internal force density opposing the external forces. The surface integral gives rise to a separate Euler Lagrange equation balancing the traction force $\sigma \mathbf{n} $ against applied surface loads.  Traction forces will not be considered here and we have left them out.  Let us reiterate one more that all of this is standard continuum mechanics\cite{Ogden1997}. We have noted it down here for later reference. 
 
 Returning to the Piola stress for a simple liquid Eq.~\ref{eqn:Piolasr} and applying the ``push forward'' transformation Eq.~\ref{eqn:Piola} we see that the Cauchy stress tensor generated by the short range interactions must be
\begin{equation}
\boldsymbol{\sigma}_{sr} =
\left(  - \rho\frac{\partial \phi}{\partial \rho}  + \phi \right)  \mathbf{I} = -P_{sr} \mathbf{I}
\label{eqn:Cstrsr}
\end{equation}
 The second identity defines the mechanical short range pressure $P_{sr}$ which is indeed equal to the local  thermodynamic pressure as obtained from variation in the density in section \ref{sec:chemech}. Clearly, without the complication of electrostatics,  the DFT and the continuum mechanics of a simple fluid are consistent. The question is whether this remains true when dielectric interactions are included in the stress tensor. 

\section{Deformation in the simple dielectric fluid } \label{sec:varF}

\subsection{Electric field stress tensor} \label{sec:estress}
To determine the variation of dielectric energy with geometry we  expand in the first order difference of the various geometry dependent quantities, similar to Eq.~\ref{eqn:dResr}, while keeping the \emph{reference frame} vector potential $\hat{\mathbf{A}}$ and polarization $\mathbf{P}$  fixed. We begin with the external field interaction energy Eq.~\ref{eqn:teI}.
\begin{equation}
  \delta\tilde{\mathcal{E}}_I   =  - \int_{\Omega_r} \mathbf{P} \cdot  \delta \mathbf{E}_0 \, dV
  = - \int_{\Omega_r}  \mathbf{P} \cdot \left( \delta \mathbf{F}^{\textrm{T}} \mathbf{e}_0 
     + \mathbf{F}^{\textrm{T}}  \delta \mathbf{e}_0  \right) dV
 \label{eqn:dteI}
\end{equation}
$ \delta \mathbf{e}_0$ is finite for-uniform applied fields. In spatial coordinates this would be simply $\delta \mathbf{e}_0 = \grad \mathbf{e}_0  \cdot\mathbf{u}$ where $\mathbf{u}$ is the first order variation  in the placement $\mathbf{x}(\mathbf{X})$ defined in Eq.~\ref{eqn:xdv}. Being directly proportional to $\mathbf{u}$ the  $ \delta \mathbf{e}_0$  term in Eq.~\ref{eqn:dteI} acts, not as a stress, but as a body force (It was this distinction that led to the Cauchy equation Eq.~\ref{eqn:Cauchy} as explained in  section \ref{sec:varclas}).  Reverting back  to  current space using Eq.~\ref{eqn:RP} we can write
 \begin{equation*}
 \int_{\Omega_r}  \mathbf{P} \cdot \left(\mathbf{F}^{\textrm{T}}  \delta \mathbf{e}_0  \right)dV 
  =  \int_{\Omega_r} \left( J^{-1} \mathbf{F}  \mathbf{P}  \right) \cdot
   \left( \grad \mathbf{e}_0 \cdot \mathbf u\right) J dV
    =    \int_{\Omega}  \mathbf{f}_I \cdot \mathbf{u} \, dv
 \end{equation*}
 with $\mathbf{f}_I$  the  Kelvin force due to a gradient in the applied field as defined in Eq.~\ref{eqn:fI}. As pointed out  in section \ref{sec:varclas} care must be taken with  sign when introducing a body force.   A displacement in response to a positive body force decreases the energy of a system as implied by  Eq.~\ref{eqn:bwork}.  Comparing to Eq.~\ref{eqn:dteI}  we see that the sign of the Kelvin force of Eq.~\ref{eqn:fI} is  consistent with that for a body force. 

 The  $\delta \mathbf{F}^{\textrm{T}}$ term  in Eq.~\ref{eqn:dteI} defines a genuine stress which is made explicit by rewriting the integral as
\begin{eqnarray*}
  \int_{\Omega_r}  \mathbf{P} \cdot 
  \left(\delta \mathbf{F}^{\textrm{T}} \mathbf{e}_0 \right) dV & =  &\int_{\Omega_r} 
      J \mathbf{p} \cdot \left( \mathbf{F}^{-\textrm{T}} \delta \mathbf{F}^{\textrm{T}} 
       \mathbf{e}_0 \right) dV
  \\ & = &      \int_{\Omega_r}  \left( J \left( \mathbf{p} \otimes \mathbf{e}_0 \right) 
  \mathbf{F}^{-\textrm{T}} \right) : \delta \mathbf{F} dV
  \end{eqnarray*}
which is of the  form of Eq.~\ref{eqn:Pwork} with a Piola stress tensor
\begin{equation*}
\boldsymbol{\Sigma}_I =  J \left( \mathbf{p} \otimes \mathbf{e}_0 \right)  \mathbf{F}^{-\textrm{T}} 
\end{equation*}
The corresponding Cauchy stress is according to Eq.~\ref{eqn:Piola}
\begin{equation}
\boldsymbol{\sigma}_I = J^{-1} \boldsymbol{\Sigma}_I\mathbf{F}^{\textrm{T}} =
\left( \mathbf{p} \otimes \mathbf{e}_0 \right)
\label{eqn:CstressI}
\end{equation}
Note that $\boldsymbol{\sigma}_I $ is asymmetric ($\mathbf{p} \otimes \mathbf{e}_0  \neq
\mathbf{e}_0\otimes \mathbf{p} $).

The procedure for extracting a stress tensor from the differential of the self energy Eq.~\ref{eqn:dpDP} is similar if more involved. There two terms due to variation of $J^{-1}$ and  $\mathbf{C}$
 \begin{eqnarray*}
\delta \tilde{\mathcal{E}}_S & = & \int_V \left(
 \frac{\delta J^{-1}}{2\epsilon_0} \left( \Curl \hat{\mathbf{A}} - \mathbf{P} \right) \cdot
\mathbf{C}\left( \Curl \hat{\mathbf{A}} - \mathbf{P} \right) \right.
\\ & & \left. + \frac{J^{-1}}{2\epsilon_0} 
\left( \Curl \hat{\mathbf{A}} - \mathbf{P} \right) \cdot
\delta \mathbf{C} \left( \Curl \hat{\mathbf{A}} - \mathbf{P} \right)
 \right) dV
\end{eqnarray*}
Substituting Eqs.~\ref{eqn:dRJ} and \ref{eqn:dC} we obtain
\begin{eqnarray*}
\delta \tilde{\mathcal{E}}_S & = & \int_V \left( 
- \frac{ J^{-1}}{2\epsilon_0} \left( \Curl \hat{\mathbf{A}} - \mathbf{P} \right) \cdot
\mathbf{C}\left( \Curl \hat{\mathbf{A}} - \mathbf{P} \right)
\mathrm{Tr} \left( \mathbf{F}^{-1} \delta \mathbf{F} \right) \right.
\\ & & \left. + \frac{J^{-1}}{2\epsilon_0} 
\left( \Curl \hat{\mathbf{A}} - \mathbf{P} \right) \cdot
\left( \delta  \mathbf{F}^{\mathrm{T}} \mathbf{F} +
         \mathbf{F}^{\mathrm{T}} \delta  \mathbf{F} \right) 
 \left( \Curl \hat{\mathbf{A}} - \mathbf{P} \right)
 \right) dV
\end{eqnarray*}
Next reinserting the Lagrangian electric field using  Eqs.~\ref{eqn:DcurlA} and \ref{eqn:RhatLorentz} and splitting the $\mathbf{C}^{-1}= \mathbf{F}^{-1} \mathbf{F}^{-\textrm{T}}$ matrix changes this to an expression  ready to be transformed back to the  current frame applying the inverse of Eqs.~\ref{eqn:RhatE}
and \ref{eqn:RP}
\begin{eqnarray*}
\delta \tilde{\mathcal{E}}_S & = & \int_V J \left( - \frac{ \epsilon_0}{2}
 \left(\mathbf{F}^{-\mathrm{T}} \hat{\mathbf{E}}\right) \cdot 
 \left(\mathbf{F}^{-\mathrm{T}} \hat{\mathbf{E}} \right)
\mathrm{Tr} \left( \mathbf{F}^{-1} \delta \mathbf{F} \right) \right.
\\ & & \left. +  \frac{ \epsilon_0 }{2}
 \left( \mathbf{F}^{-\textrm{T}} \hat{\mathbf{E}} \right) \cdot
 \mathbf{F}^{-\mathrm{T}}  \left( \delta  \mathbf{F}^{\mathrm{T}} \mathbf{F} +
         \mathbf{F}^{\mathrm{T}} \delta  \mathbf{F} \right) \mathbf{F}^{-1}
\left(  \mathbf{F}^{-\textrm{T}} \hat{\mathbf{E}} \right) \right) dV
 \\ & = &
\int_V  J \left( - \frac{ \epsilon_0}{2} \left( \hat{\mathbf{e}}\cdot \hat{\mathbf{e}} \right)
\mathrm{Tr} \left( \mathbf{F}^{-1} \delta \mathbf{F} \right) 
   +  \frac{ \epsilon_0}{2} \hat{\mathbf{e}} \cdot 
\left( \mathbf{F}^{-\mathrm{T}} \delta  \mathbf{F}^{\mathrm{T}}
 + \delta  \mathbf{F} \, \mathbf{F}^{-1} \right) \hat{\mathbf{e}} \right) dV
\label{eqn:defdF}
\end{eqnarray*} 
This is then recast in the form of Eq.~\ref{eqn:Pwork} 
\begin{equation}
 \delta \tilde{\mathcal{E}}_S = \int_V  J \left(  - \frac{ \epsilon_0}{2}
 \left( \hat{\mathbf{e}}\cdot \hat{\mathbf{e}} \right) \mathbf{F}^{-\textrm{T}} 
 +  \epsilon_0 \left( \hat{\mathbf{e}} \otimes \hat{\mathbf{e}} \right)
\mathbf{F}^{-\textrm{T}} \right): \delta \mathbf{F} dV
\label{eqn:dePiola}
\end{equation}
defining the Piola-Maxwell stress tensor\cite{Suo2008,Liu2014}
\begin{equation}
\hat{\boldsymbol{\Sigma}}_{\textrm{M}} = J \left(
 \epsilon_0 \left( \hat{\mathbf{e}} \otimes \hat{\mathbf{e}} \right) 
- \frac{ \epsilon_0}{2} \left( \hat{\mathbf{e}}\cdot \hat{\mathbf{e}} \right) \mathbf{I}
  \right)\mathbf{F}^{-\textrm{T}}
\label{eqn:MPstress}
\end{equation}
We left a hat on $\hat{\boldsymbol{\Sigma}}_{\textrm{M}}$ is a reminder  that only self fields contribute. Applying transformation Eq.~\ref{eqn:Piola} we obtain the ``self'' Maxwell stress tensor in the current frame
\begin{equation}
\hat{\boldsymbol{\sigma}}_{\textrm{M}}  = \epsilon_0
\left(\hat{\mathbf{e}} \otimes \hat{\mathbf{e}} \right) - \frac{ \epsilon_0}{2}
 \left( \hat{\mathbf{e}}\cdot \hat{\mathbf{e}} \right) \mathbf{I}
\label{eqn:MCstress}
\end{equation}
Combining with the external field interaction stress Eq.~\ref{eqn:CstressI} (taking into account the minus sign in Eq.~\ref{eqn:dteI})   yields a stress tensor
 \begin{equation}
\boldsymbol{\sigma}_{\textrm{E}} =
\hat{\boldsymbol{\sigma}}_{\textrm{M}}  -  \boldsymbol{\sigma}_I =
\epsilon_0\left(\hat{\mathbf{e}} \otimes \hat{\mathbf{e}} \right)  
 - \left( \mathbf{p} \otimes \mathbf{e}_0 \right)
- \frac{ \epsilon_0}{2} \left( \hat{\mathbf{e}}\cdot \hat{\mathbf{e}} \right) \mathbf{I}
 \label{eqn:Ekstress}
 \end{equation}
 $\boldsymbol{\sigma}_{\textrm{E}}$ can be interpreted as the stress in response to deformation of the Ericksen field energy Eq.~\ref{eqn:Ediel}.  However, keep in mind that this also produced a body force Eq.~\ref{eqn:fI}.

\subsection{Polarization stress tensor} \label{sec:dostress}
Variation of  $\tilde{\mathcal{F}}_P$  is partly similar  to the procedure for the electrostatic field energy leading to Eq.~\ref{eqn:MPstress}. The reason is that the pull back rule for polarization(Eq.~\ref{eqn:RP}) is the same as for the dielectric displacement field (Eq.~\ref{eqn:RD}). This was in fact already used in Eq.~\ref{eqn:dpDP}. Going through the same steps that led to Eq.~\ref{eqn:defdF} we start off with
\begin{eqnarray*}
\delta \tilde{\mathcal{F}}_P & = & 
 \int_V \left(  \frac{ \delta J^{-1}}{2 \chi}  
- \frac{J^{-1}}{2 \chi^2} \left(\frac{\partial \chi}{\partial \rho}
\right) \delta \rho \right)
 \mathbf{P} \cdot \left( \mathbf{C} \mathbf{P} \right) dV
\\ & & + \int_V  \frac{J^{-1}}{2 \chi} 
\mathbf{P} \cdot \left( \delta \mathbf{C} \mathbf{P} \right) dV
\end{eqnarray*}
Continuing as we did for the  self field energy we apply Eq.~\ref{eqn:dRrho} followed by Eq.~\ref{eqn:dRJ} to convert $\delta \rho$ and $\delta J$ and find 
\begin{eqnarray*}
\delta \tilde{\mathcal{F}}_P   & = & 
 \int_V J \left( -\frac{1}{2 \chi} + \frac{\rho}{2 \chi^2}
 \left(\frac{\partial \chi}{\partial \rho}\right) \right) \mathbf{p}^2
\mathrm{Tr}  \left(\mathbf{F}^{-1} \delta \mathbf{F} \right)  dV
\\ & & + \int_V  \frac{J}{2 \chi} \mathbf{p}
 \cdot \left( \mathbf{F}^{-\mathrm{T}} \delta  \mathbf{F}^{\mathrm{T}} +
  \delta  \mathbf{F}\mathbf{F}^{-1} \right) \mathbf{p} dV
\end{eqnarray*}
Factoring out the variation in the deformation gradient gives
\begin{equation*}
\delta \tilde{\mathcal{F}}_P  =  
\int_V J \left( -\frac{\mathbf{p}^2}{2 \chi} \mathbf{F}^{-\textrm{T}}+ \frac{\rho}{2 }
 \left(\frac{\partial \chi}{\partial \rho}\right)  \frac{\mathbf{p}^2}{ \chi^2}\mathbf{F}^{-\textrm{T}}
+  \frac{1}{\chi} \left( \mathbf{p} \otimes \mathbf{p} \right) \mathbf{F}^{-\textrm{T}}
 \right) : \delta \mathbf{F}  dV
\end{equation*}
resulting in a  Piola polarization stress tensor
\begin{equation}
\boldsymbol{\Sigma}_P =  J \left( \frac{1}{\chi}
 \left( \mathbf{p} \otimes \mathbf{p} \right)
 + \left( -\frac{\mathbf{p}^2}{2 \chi} + \frac{\rho}{2 }
 \left(\frac{\partial \chi}{\partial \rho}\right) \frac{\mathbf{p}^2}{\chi^2} \right) \mathbf{I}
 \right)\mathbf{F}^{-\textrm{T}}
\label{eqn:PPstress}
\end{equation}
Using once more the stress tensor transformation rule Eq.~\ref{eqn:Piola} we obtain the Cauchy polarization stress tensor 
\begin{equation}
\boldsymbol{\sigma}_P  = \boldsymbol{\sigma}_{\textrm{DO}} + \frac{\rho}{2 \chi^2}
 \left(\frac{\partial \chi}{\partial \rho}\right)  \mathbf{p}^2 \mathbf{I}
 \label{eqn:PDOstress}
\end{equation}
with $\boldsymbol{\sigma}_{\textrm{DO}}$ given by
\begin{equation}
\boldsymbol{\sigma}_{\textrm{DO}} = \frac{1}{\chi} \left( \mathbf{p} \otimes \mathbf{p} \right)  
   - \frac{1}{2 \chi} \left(\mathbf{p} \cdot \mathbf{p} \right)  \mathbf{I}
   \label{eqn:DOstress}
\end{equation}
 Note the similarity between $\boldsymbol{\sigma}_{\textrm{DO}}$  and the Maxwell stress tensor $\hat{\boldsymbol{\sigma}}_{\textrm{M}}$ of Eq.~\ref{eqn:MCstress}.  Both tensors are symmetric.   Substituting the  constitutive relation Eq.~\ref{eqn:pchie} converts $\boldsymbol{\sigma}_P$ to the seemingly asymmetric form 
 \begin{equation}
\boldsymbol{\sigma}_P =\left( \mathbf{p} \otimes \mathbf{e} \right)  
   - \frac{1}{2} \left(\mathbf{p} \cdot \mathbf{e} \right)  \mathbf{I}+ \frac{\rho}{2}
 \left(\frac{\partial \chi}{\partial \rho}\right)  \mathbf{e}^2 \mathbf{I}
\label{eqn:PCstress}
\end{equation}
 
 The appearance of a dyadic product  term  in stress due to stored polarization energy may be somewhat unexpected. Such a contribution, which is the hallmark of shape dependence, is in fact absent in the derivations of Refs.~\citenum{Trianta2018} and \citenum{Liu2014} treating polarization as a density rather than a displacement field.  The latter option is the special feature  of the  Dorfmann-Ogden scheme used here (Eq.~\ref{eqn:RP}).   $\boldsymbol{\sigma}_{\textrm{DO}}$ of Eq.~\ref{eqn:DOstress} will therefore be referred to as the Dorfmann-Ogden  stress tensor. 
 
 \subsection{Total dielectric stress tensor and force density} \label{sec:elemech}
 The various partial stresses determined in the previous section are now assembled in a total stress. Superimposing the three dyadic product terms in Eqs.~\ref{eqn:Ekstress} and \ref{eqn:PCstress}  using the field relations  Eqs.~\ref{eqn:eMax} and \ref{eqn:hatLorentz} we find they can be collapsed in a single tensor product
\begin{equation}
\epsilon_0 \left( \hat{\mathbf{e}} \otimes \hat{\mathbf{e}} \right) 
- \left( \mathbf{p} \otimes \mathbf{e}_0 \right)  +\left( \mathbf{p} \otimes \mathbf{e} \right)=
 \hat{\mathbf{d}} \otimes \hat{\mathbf{e}}
 \label{eqn:dcrosse}
\end{equation}
  Adding the isotropic term of the Maxwell stress tensor Eq.~\ref{eqn:MCstress}  yields the Toupin stress tensor $\boldsymbol{\sigma}_{\mathrm{T}}$  of Eq.~\ref{eqn:lhatstress}.  The final step is to include the  isotropic component of the polarization stress tensor of Eq.~\ref{eqn:PCstress} and the short range interactions (Eq.~\ref{eqn:Cstrsr}). The result is $\boldsymbol{\sigma}_d$ of Eq.~\ref{eqn:sigmad}. This is the formulation of  dielectric stress tensor as given by Ericksen\cite{Ericksen2007}. Alternatively $\sigma_d$ can be  expressed in terms of the stress tensors of Eq.~\ref{eqn:Ekstress} and  Eq.~\ref{eqn:DOstress} using the relation
\begin{equation}
\hat{\boldsymbol{\sigma}}_{\mathrm{T}} - \frac{1}{2} \left(\mathbf{p} \cdot \mathbf{e} \right) \mathbf{I}  =  {\boldsymbol{\sigma}}_{\mathrm{E}}  +
 \boldsymbol{\sigma}_{\mathrm{DO}}  
\label{eqn:TEDO}
\end{equation}
The virtue of this decomposition is explicit separation between electrostatic field stress  ${\boldsymbol{\sigma}}_{\mathrm{E}}$  and Dorfmann-Ogden stress $\boldsymbol{\sigma}_{\mathrm{DO}} $, which is of constitutive origin.

We have now come full circle and are ready to go  back to the question of the force density of section \ref{sec:chemech}.  This amounts to determining  the divergence of $\boldsymbol{\sigma}_{d} $. 
We will do this for $\boldsymbol{\sigma}_{d} $ of Eq.~\ref{eqn:sigmad}.    To find a convenient expression for the divergence of the dyadic product we  first  investigate the force density due to the Maxwell stress tensor $\boldsymbol{\hat{\sigma}}_{\textrm{M}}$ of Eq.~\ref{eqn:MCstress}. The divergence of the dyadic product in Eq.~\ref{eqn:MCstress} gives according to Eq.~\ref{eqn:divutimesv}
\begin{equation}
 \ddiv \epsilon_0 \left(  \hat{\mathbf{e}} \otimes \hat{\mathbf{e}} \right) = 
 \epsilon_0 \hat{\mathbf{e}} \, \ddiv \hat{\mathbf{e}}  + 
\epsilon_0 \left( \hat{\mathbf{e}} \cdot  \grad \right) \hat{\mathbf{e}}
\label{eqn:divexe}
\end{equation}
Then there still is the isotropic component  in $\boldsymbol{\hat{\sigma}}_{\textrm{M}}$. The divergence of the inproduct  of the self field with itself is worked out   by  first applying Eq.~\ref{eqn:divphiI} followed by Eq.~\ref{eqn:divuinv}. Because  $\curl \hat{\mathbf{e}} = 0$ the gradient of $\left( \hat{\mathbf{e}} \cdot \hat{\mathbf{e}} \right)$  reduces to just one term 
 \begin{equation}
  \ddiv \frac{\epsilon_0}{2} \left( \hat{\mathbf{e}} \cdot \hat{\mathbf{e}} \right) \mathbf{I} =
\frac{\epsilon_0}{2} \grad \left( \hat{\mathbf{e}} \cdot \hat{\mathbf{e}} \right) =
 \epsilon_0 \left( \hat{\mathbf{e}} \cdot \grad \right) \hat{\mathbf{e}}
\label{eqn:diveine}
\end{equation}
which cancels against the same term in Eq.~\ref{eqn:divexe}. The result is the self force acting on the response charge density $\hat{q}$
\begin{equation}
 \ddiv \hat{\boldsymbol{\sigma}}_{\textrm{M}} = 
 \epsilon_0 \hat{\mathbf{e}} \, \ddiv \hat{\mathbf{e}} = \hat{q} \hat{\mathbf{e}}
\label{eqn:qforce}
\end{equation} 
Recall that for the pure dielectric $\hat{q} = q$  is the total charge density. 

The difference between  $\boldsymbol{\sigma}_{\textrm{M}}$  and the Toupin stress tensor  $\boldsymbol{\sigma}_{\textrm{T}}$  (Eq.~\ref{eqn:lhatstress}) is that  $\epsilon_0 \hat{\mathbf{e}} \otimes \hat{\mathbf{e}}$ is replaced by  $\hat{\mathbf{d}} \otimes \hat{\mathbf{e}} $.  Using Eq.~\ref{eqn:divutimesv}, setting $\ddiv \mathbf{d} = 0 $ and substituting  Eq.~\ref{eqn:hatLorentz} yields
\begin{equation*}
 \ddiv \left(  \hat{\mathbf{d}} \otimes \hat{\mathbf{e}} \right) 
 =  \left( \mathbf{p} \cdot  \grad \right) \hat{\mathbf{e}} +
 \epsilon_0 \left( \hat{\mathbf{e}} \cdot  \grad \right) \hat{\mathbf{e}}
\end{equation*}  
The divergence of the inproduct term is still given by Eq.~\ref{eqn:diveine}.  The same cancellation as for $\ddiv \boldsymbol{\sigma}_{\textrm{M}}$ leads to
\begin{equation}
 \ddiv \hat{\boldsymbol{\sigma}}_{\textrm{T}} 
 = \left( \mathbf{p} \cdot \grad \right) \hat{\mathbf{e}}
\label{eqn:pforce}
\end{equation} 
The Lorentz force  Eq.~\ref{eqn:qforce}  on the scalar polarization charge has become a  Kelvin force acting on  vectorial polarization.  

What to do with the inproduct $\left(\mathbf{p} \cdot \mathbf{e} \right)$ in Eq.~\ref{eqn:sigmad}? It is an isotropic stress so we could leave it as the gradient of a pressure. However for the purpose of comparison  to the DFT force density we follow Haus and Melcher (Ref.~\citenum{Melcher1989}, chapter 11) and write $(\mathbf{p}\cdot \mathbf{e})$ as $ \chi (\mathbf{e}\cdot \mathbf{e})$ and apply the chain rule treating  $(\mathbf{e}\cdot \mathbf{e}) = \mathbf{e}^2$ as a scalar. This gives 
\begin{equation*}
\grad (\mathbf{p}\cdot \mathbf{e}) = \chi \,\grad (\mathbf{e}\cdot \mathbf{e}) +
   \mathbf{e}^2 \grad \chi
\end{equation*}
Next we use Eq.~\ref{eqn:divuinv} on $(\mathbf{e}\cdot \mathbf{e})$. Because $\curl \mathbf{e} = 0 $ the result is
\begin{equation*}
\grad (\mathbf{e}\cdot \mathbf{e}) = 2 \left(\mathbf{e} \cdot \grad \right)\mathbf{e}
\end{equation*}
Combining these two expressions once more using $\mathbf{p} = \chi \mathbf{e}$ we arrive at
\begin{equation*}
\grad (\mathbf{p}\cdot \mathbf{e}) =  2 \left(\mathbf{p}\cdot \grad \right) \mathbf{e} +
\mathbf{e}^2 \grad \chi
\end{equation*}
Substituting this together with  Eq.~\ref{eqn:pforce} yields an internal force density
\begin{equation*}
\ddiv \boldsymbol{\sigma}_{\textrm{d}} =  \left( \mathbf{p} \cdot \grad \right) \hat{\mathbf{e}} -
\left(\mathbf{p}\cdot \grad \right) \mathbf{e} -\frac{\mathbf{e}^2}{2} \grad \chi  
+ \grad \left( \frac{\rho}{2} \left(\frac{\partial \chi}{\partial \rho}\right)  \mathbf{e}^2   - P_{sr}\right)
\end{equation*}
With  Eq.~\ref{eqn:eMax} only the gradient of the external field remains.  In the last three terms we recover the Korteweg-Helmholtz force density  $\mathbf{f}_{\textrm{KH}}$ of Eq.~\ref{eqn:fKH} leading to Eq.~\ref{eqn:dforce} of section \ref{sec:chemech}.

We have now arrived at a subtle point, the most surprising of the many cancellations we have already encountered.  The Kelvin force  $\mathbf{f}_I$  of Eq.~\ref{eqn:fI} makes a double appearance, as  an electrical  body force  in Eq.~\ref{eqn:dteI} and a second time in  Eq.~\ref{eqn:dforce} where, in the language of section \ref{sec:varclas}, it acts as an internal force resisting perturbation by external forces.   This means that setting up the Cauchy equation (Eq.~\ref{eqn:Cauchy}) for the polarized fluid $\mathbf{f}_I$ will have to be included as a further body force, which explains the $\mathbf{f}_I$ in Eq.~\ref{eqn:dCauchy}. The sign is crucial.  The argument starting off section \ref{sec:estress} was meant to show that $\mathbf{f}_I$, in its role as body force must be added with a positive sign to the force balance Eq.~\ref{eqn:dCauchy} with the consequence  that it cancels the internal Kelvin force in Eq.~\ref{eqn:dforce}.  $\mathbf{f}_I$ has been eliminated. The Cauchy equilibrium condition Eq.~\ref{eqn:dCauchy} is reduced to $\mathbf{f}_{\textrm{KH}} + \mathbf{f}_{ext} = 0 $. After a huge effort we have arrived back at the DFT equilibrium equation. There must be a simple physical argument to rationalize or even predict this outcome. Such an explanation  escapes us at this stage. We can however point out that  cancellation of the external Kelvin force is not a consequence  of the linear constitutive relation of the simple dielectric fluid.  It is a property of the Ericksen field energy Eq.~\ref{eqn:Ekel} and will therefore occur whenever this very general form of electrostatic field energy is used.

\section{Discussion and summary} \label{sec:disc}
The paper presents a  detailed comparison between a density functional style and continuum mechanics treatment of the simplest of continuum dielectric energy functionals  (the Marcus-Felderhof functional).  The motivation for this study is  the observation that dielectric  energy is sensitive to changes in configuration not covered by variation in  local density. This additional configurational degree of freedom (``kinematic descriptor'') is volume conserving deformation.  This suggested to us that  the DFT equilibrium equation, formulated as a force balance, could be missing terms related to the response to deformation.  We found that this is not the case due to a somewhat mysterious series of cancellations.  The central tool in this derivation is a variational method introduced  by Ericksen using  a vector potential as an additional (extended) variational degree of freedom\cite{Ericksen2007}. The function of the vector potential is to generate the divergence free dielectric displacement characteristic of pure dielectrics.   The variational procedure  was implemented using the Lagrangian approach to electro-elasticity developed by Dorfmann and Ogden\cite{Ogden2005,Ogden2009}.  The pivotal equations in this formalism are the Lagrangian (material)  representations of the fields ($\mathbf{e}, \mathbf{d}, \mathbf{p}$) in Maxwell theory. These transformation rules capture the geometry dependence of the dielectric energy which is hidden in purely spatial (Eulerian) theory.  

The key result  is an expression for the stress tensor of the simple dielectric fluid (Eq.~\ref{eqn:sigmad}).  This expression  was obtained using a particular form of the electrostatic energy valid for finite bodies\cite{Ericksen2007}.   Finite body geometry is fundamental in continuum solid mechanics.  It creates the sharp surfaces where the environment acts on the body by way of traction forces. Realistic surfaces and interfaces in liquids are of a continuous (diffuse) nature as clearly demonstrated by the classical DFT of inhomogeneous liquids\cite{Evans1979}.  We have retained a finite body geometry in this first application  to  dielectric liquids, because it  allows for a consistent separation between the Maxwell stress in an empty container and the stress induced in the dielectric. Boundary and jump conditions were completely ignored in this presentation.  We have deferred this important issue as a subject of future research where also the question of diffuse surfaces will have to be addressed.

The quantification of stress in deformable dielectric continua remains subject to some debate\cite{Kovetz2000,Brenner2002,Landis2005,Ericksen2007b,Steigmann2009,Bustamante2009}. A thermomechanically consistent  theory most likely requires a non-equilibrium  electrodynamics framework considering fluxes and energy balance equations.  The work reported here makes no claim to have contributed to resolving these issues.  It is intended as a test  of the consistency between a coupled density functional type approach and continuum mechanics applied to exactly the same elementary non-equilibrium polarization energy  functional.  In this respect it should be mentioned that there was also a more practical inspiration  for the work presented here, namely the results of a recent paper with Chao Zhang\cite{Zhang2020}.   There we applied finite field molecular dynamics simulation\cite{Zhang2020b} to study the electromechanical properties of the water liquid-vapour interface. We computed the stress tensor using a standard SPC force field and found that there is a qualitative difference in response to a perpendicular applied field depending on the proximity to the planar interface. Using a phenomenological electromechanical model this observation was interpreted as evidence for the role of phoretic forces generated by the inhomogeneity of the Maxwell field in the interface layer.  However the electromechanical theory was very heuristic  requiring rigorous justification and probably also adjustments.  The derivation in this paper is a preparation for this task. 

We end with a speculative comment. This concerns possible application of continuum mechanics based methods  to  molecular density functional theory (MDFT) of polar fluids. MDFT is a proper density functional theory based on the joint  singlet position orientation density\cite{TelodaGama1991,Gray1992,Gray1994,Oxtoby1993,Dietrich1992,Dietrich1993,Dietrich1994,Dietrich1996,Warschavsky2002,Warschavsky2003,Borgis2012,Borgis2013,Borgis2017}.  As such it is a promising tool for study of the electromechanics of polar fluids capable of reaching down to microscopic length scales inaccessible to  the constitutive models used in continuum mechanics.  As in the original atomistic Hamiltonian, orientations in MDFT are coupled by the  dipole-dipole interaction. This is of course is a notoriously dangerous long-range interaction and the design of approximate energy functionals for dipolar fluids requires careful consideration\cite{TelodaGama1991,Dietrich1992}  even more so than for ionic liquids\cite{Evans1980,Hansen2000,Haertel2017}. A representation in terms of electric fields could put these problems in a different perspective closer to the  profound principles of Maxwell theory. The simple dielectric fluid model of section \ref{sec:Fdiel} is a basic example of such  a Maxwell-Lorentz  enhanced functional.  The prevalence  of Lagrangian methods in non-linear continuum mechanics is moreover an argument to try to extrapolate these methods to the density functional theory of molecular liquids for problems where knowledge of the stress tensor is required.   Despite the extra burden of the vector and tensor analysis such a material field based  approach  may also offer computational advantages.  We hope that our paper can be an encouragement for  MDFT experts to look into this challenging possibility.

\section*{Acknowledgement}

Stephen Cox and Robert Evans  are acknowledged for helpful discussions on this challenging subject. 
\appendix
\renewcommand{\theequation}{A.\arabic{equation}}
\setcounter{equation}{0}

\section{Vector identities}

$\mathbf{u}$ and $\mathbf{v}$ are vector fields $\mathbf{f}(\mathbf{r}),\mathbf{g}(\mathbf{r})$. $\phi$ is a scalar function $\phi(\mathbf{r})$ 
\begin{gather}
\ddiv\left( \mathbf{u} \times \mathbf{v} \right) = \mathbf{v} \cdot \left( \curl \mathbf{u} \right) - \mathbf{u}  \cdot \left( \curl \mathbf{v} \right)
\label{eqn:divuoutv} \\
\curl \curl \mathbf{u} = \grad \ddiv \mathbf{u} - \Delta \mathbf{u}
\label{eqn:curl2} \\
\ddiv \left( \phi \mathbf{I} \right) = \grad \phi
\label{eqn:divphiI} \\
\grad \left( \mathbf{u} \cdot \mathbf{v} \right ) =
\left( \mathbf{u} \cdot \grad \right) \mathbf{v} + 
\left( \mathbf{v} \cdot \grad \right) \mathbf{u} +
\mathbf{u} \times \curl \mathbf{v} + \mathbf{v} \times \curl \mathbf{u}
\label{eqn:divuinv} \\
\ddiv \left( \mathbf{u} \otimes \mathbf{v} \right) = 
\left( \ddiv \mathbf{u} \right) \mathbf{v} + 
\left( \mathbf{u} \cdot \grad\right) \mathbf{v}
\label{eqn:divutimesv}  
\end{gather}


\end{document}